\documentclass{aastex}
\usepackage{epsfig}
 \usepackage{graphicx}
\shorttitle{Hubble flow around Cen A/M 83 }
\shortauthors{Karachentsev et al.}

\begin{document}
\title{The Hubble flow around the Cen A/M 83 galaxy complex }

\author{Igor D.\ Karachentsev}
\affil{Special Astrophysical Observatory, Russian Academy
	  of Sciences, N.\ Arkhyz, KChR, 369167, Russia}
\email{ikar@luna.sao.ru}
\author{R. Brent Tully}
\affil{Institute for Astronomy, University of Hawaii, Honolulu, HI 96822}
\author{Andrew Dolphin}
\affil{Steward Observatory, 933 N. Cherry Ave., Tucson, AZ 85721}
\author{Margarita Sharina, Lidia Makarova\altaffilmark{1,2} and Dmitry Makarov\altaffilmark{1,2}}
\affil{Special Astrophysical Observatory, Russian Academy
	  of Sciences, N.\ Arkhyz, KChR, 369167, Russia}
\author{Shoko Sakai}
\affil{Division of Astronomy and Astrophysics, University of California,
       at Los Angeles, Los Angeles, CA 90095-1562}
\author{Edward J. Shaya}
\affil{Astronomy Department, University of Maryland, College Park, MD 20743}
\author{Olga G. Kashibadze}
\affil{Moscow State University}
\author{Valentina Karachentseva}
\affil{Astronomical Observatory of Kiev University, Kiev 04053, Ukraine}
\author{Luca Rizzi}
\affil{Institute for Astronomy, University of Hawaii, Honolulu, HI 96822}

\altaffiltext{1}{also Institute for Astronomy, University of Hawaii,
2680 Woodlawn Drive, Honolulu, HI 96822}
\altaffiltext{2}{Isaac Newton Institute of Chile, SAO Branch}

\begin{abstract}
 We present HST/ACS images and color-magnitude diagrams for 24 nearby
galaxies in and near the constellation of Centaurus with radial velocities 
$V_{LG}< 550$ km s$^{-1}$.
Distances are determined based on the luminosities of stars at the
tip of the red giant branch that range from 3.0~Mpc to 6.5~Mpc.
The galaxies are concentrated in two spatially separated groups around
Cen~A (NGC~5128) and M~83 (NGC~5236).  The Cen~A group itself has a mean
distance of 3.76$\pm$0.05 Mpc, a velocity dispersion of 136 kms$^{-1}$,
a mean harmonic radius of 192 kpc, and an estimated orbital/virial mass of
{\bf $(6.4 - 8.1)\cdot 10^{12} M_{\sun}$.}  This elliptical dominated group
is found to have a relatively high mass-to-light ratio: 
$M/L_B = 125~M_{\odot}/L_{\odot}$.
For the M~83 group we derived a mean
distance of 4.79$\pm$0.10 Mpc, a velocity dispersion of 61 km s$^{-1}$,
a mean harmonic radius of 89 kpc, and estimated orbital/virial mass of
$(0.8 - 0.9)\cdot 10^{12} M_{\sun}$.  This spiral dominated group is
found to have a relatively low $M/L_B = 34~M_{\odot}/L_{\odot}$. 
The radius of the zero-velocity surface around
Cen~A lies at
 $R_0 = 1.40\pm0.11$ Mpc. implying a total mass within $R_0$ of 
$M_T = (6.0\pm1.4) \cdot 10^{12} M_{\sun}$.  This value is in good
agreement with the Cen~A virial/orbital mass
estimates and provides confirmation of the relatively high $M/L_B$ of 
this elliptical--dominated group.  The centroids of both the groups, 
as well as
surrounding field galaxies, have very small peculiar velocities,
$ < 25$ km s$^{-1}$, with respect to the local Hubble
flow with $H_0$ = 68 km s$^{-1}$ Mpc$^{-1}$.

\end{abstract}

\section{Introduction}

The distribution of dark versus luminous matter on scales of 0.1 -- 1.0 Mpc
still remains poorly understood.  The situation has been improving thanks to
recent observations with Hubble Space Telescope (HST) that provide images of 
the resolved stellar content of hundreds of nearby galaxies.  The brightness
of the tip of the red giant branch (TRGB) in these images provide accurate
distances.   These accurate distances combined with accurate 
observed velocities, mostly from HI observations, allow for the decoupling
of `peculiar' velocities from the cosmic expansion.  These peculiar 
velocities tell us something about the distribution of matter.

The observations of the nearest groups has been reviewed by Karachentsev 
(2005).  Masses for the groups have been determined both by traditional
measures of the internal group dynamics and by a method that uses the
observed dimensions of groups at the radii of decoupling between
group collapse and cosmic expansion.  This location between collapse and
expansion is called the ``zero--velocity surface''.  It has been found that
for the 6 nearest groups the zero--velocity surface method  gives 
mass--to--light ratios (B band) in the range [15 -- 90] $M_{\sun}/L_{\sun}$
with a median of 26 $M_{\sun}/L_{\sun}$.  If these values were universal,
they would imply a very low value for the density of matter,
$\Omega_{m,local}\sim 0.04$.
The observed ``coldness'' of the local Hubble
flow, characterized by typical random motions
$\sigma_v \sim$ 30 km s$^{-1}$ (Karachentsev et al. 2002c),
implies an extremely low value of $\Omega_m$, unless modulated by
the effects of vacuum-dominance (Chernin 2001).

Part of the discrepancy between very locally evaluated $\Omega_m$
and more global measures may be a consequence of large $M/L$
variations with environment due to astrophysical processes
like stellar aging and gas dispersal through heating
(Tully 2005; van den Bosch et al. 2003).
Although most of the light is in the ``field'', a substantial
fraction of the mass of the universe might be confined to dense
clusters dominated by early-type galaxies. Consequently one must
suitably average $M/L$ over these distinct environments
to arrive at estimates that are cosmologically meaningful.
The local volume has a low density compared with clusters, such
that in almost all the neighboring groups the
dominant members are spiral galaxies.   Bahcall et al. (2000)
suggest that giant elliptical galaxies have 
more extended and $\sim3$ times more massive halos than spiral ones. 
 
We are learning about the nature of groups in our neighborhood.
Karachentsev et al. (2002a, 2002b, 2003) have
shown that
the M~81/NGC~2403, Cen~A/M~83, and Maffei~1/IC~342 groups  are
all ``dumbbell'' systems like the Local Group. In the best studied case 
beyond the 
Local Group, the M~81/NGC~2403 Group,
the core around the foreground NGC~2403 sub-structure is clearly falling
back toward the background M~81 sub-structure.  Evidently, this group,
like the Local Group, has semi-virialized cores and a larger bound,
infalling zone.  The masses inferred for the overall bound
regions are essentially that of the sum of the cores.

The Cen~A/M~83 region is particularly interesting because one of the
dominant galaxies, Cen~A, is a giant elliptical.  The only other large
elliptical nearby is the badly obscured Maffei 1.  With Cen~A, we are 
offered the
unique opportunity to study the outer halo of a giant elliptical in the 
same way we have been studying the environments of giant spirals.
In a preliminary investigation
(Karachentsev et al. 2002b) it was already seen that there are
two distinct cores: around Cen~A to the foreground and M~83 to the
background.  The two cores have the same radial
velocities within the uncertainties of the observations so
it is not clear if the two cores are bound or escaping from
one another.  Fortunately this region is rich in dwarf galaxies.
These small galaxies provide test probes of the potential.
Here, we present TRGB distances to galaxies in the wide vicinity of the
nearby giant elliptical galaxy Cen~A to flesh out the 3D view of the
complex and improve our understanding of its dynamical state.

\section{ HST ACS photometry and Color--Magnitude diagrams}

We have observed 24 galaxies with the Advanced Camera
for Surveys (ACS) during HST Cycles 12 and 13 (proposals 9771 and 10235).
We obtained 1200s F606W and 900s F814W images of each galaxy using
ACS/WFC with exposures split to eliminate cosmic ray contamination.  
The cosmic ray cleaned images (CRJ data
sets) were obtained from the STScI archive, having been processed
according to the standard ACS pipeline.

Stellar photometry was obtained using the ACS module of DOLPHOT (Dolphin
et al. in prep), using the recommended recipe and parameters.  In brief,
this involves the following steps.  First, pixels that are flagged as bad
or saturated in the data quality images were marked in the data images.
Second, pixel area maps were applied to restore the correct count rates.
Finally, the photometry was run.
In order to be reported, a star had to be recovered with S/N of at least
five in both filters, be relatively clean of bad pixels (such that the
DOLPHOT flags are zero) in both filters, and pass our goodness of fit
criteria ($\chi \le 2.5$ and $\vert sharp \vert \le 0.3$).

To estimate our photometric uncertainties and completeness, artificial star
tests were run on the ESO~269-058 and KKs~55 fields, which represent the
most and least crowded images.  Completeness plots are shown in Figure
1.  The plateau at $\sim 85$\% completeness is due to bad
pixels and cosmic rays.  The magnitude errors as a function
of recovered magnitude are shown in Figure 2.
CTE corrections were made according to ACS ISR03-09, and our zero points
and transformations were made according to Sirianni et al. (2005).  We
estimate the uncertainties in the calibration to be around 0.05
magnitudes.

   We determined the TRGB using a Gaussian-smoothed $I$-band luminosity
function for red stars with colors $V-I$ within $\pm0\fm5$ of the mean
$\langle V-I \rangle$ expected for red giant branch stars. Following
Sakai et al. (1996), we applied a Sobel edge-detection filter.
The position of the TRGB was identified with the peak in the
filter response function. 
Uncertainty in measuring the TRGB was determined by performing bootstrap
resampling. Drawing from the original luminosity function, every stellar 
magnitude has been displaced randomly following a gaussian distribution. 
Then a new luminosity function is determined
and the tip magnitude $I_{TRGB}$ is calculated. This procedure was performed 1000 times
for each galaxy. We take the standard deviation of the distribution of $I_{TRGB}$
as the uncertainty.
According to Da Costa \& Armandroff (1990),
the TRGB is located at $M_I = -4.05$ mag.  The calibration relations were
derived over the metallicity range of the Galactic globular clusters 
$(-2.1 \leq [Fe/H] \leq 0.7)$ and are expected to work well over ages
spanning 2 -- 15 Gyr.
Ferrarese et al. (2000) calibrated the zero point
of the TRGB from galaxies with Cepheid distances and estimated
$M_I = -4\fm06 \pm 0\fm07(random)\pm0.13$(systematic). A new TRGB calibration,
$M_I = -4\fm04 \pm 0\fm12$, was made by Bellazzini et al.(2001) based on
photometry and on a distance estimate from a detached eclipsing binary in the
Galactic globular cluster $ \omega$ Centauri. For this paper (as for
our previous works with the HST data) we use $M_I = -4\fm05$.
We consider total errors in distance moduli to be the quadrature sum
of the internal errors and external systematic errors. Internal errors include
the uncertainties in the TRGB measurement, in the HST photometry zero point 
($\sim0\fm05$), the aperture corrections ($\sim0\fm05$) and the uncertainties 
in extinction ($A_I$), which are taken to be 10\% of the assumed values 
given by 
Schlegel et al. (1998). The external systematic error is the uncertainty
in the $M_{I,TRGB}$ zero-point which is taken to be $0\fm12$ following 
Bellazzini et al.(2001).

\section{TRGB distances and integrated properties of 24 galaxies}

  ACS images of 24 observed galaxies are shown in Figure 3. The compass in
each field indicates the North and East directions. Usually our target
galaxies were centered on the middle of the ACS field. In Figure 4
$I$ versus $(V-I)$ color magnitude diagrams (CMDs) for the 24 galaxies
are presented.

A summary of the resulting distance moduli for the observed galaxies
is given in Table~1. Columns contain: (1) galaxy name, (2) equatorial
coordinates, (3) radial velocity in km s$^{-1}$ in the Local Group (LG)
rest frame,
(4) apparent $I$-band magnitude of the TRGB, and (second line)
the uncertainty in measuring the TRGB,
(5) the mean $V-I$ color measured at an absolute $I$ 
magnitude $-3.5$, correlated with metallicity (Lee et al. 1993), $\pm$
rms uncertainty of the mean color, and (second line) the standard deviation
of the RGB color,
(6) Galactic extinction in the $I$-band from Schlegel et al. (1998),
(7) true distance modulus in mag and total error in the distance modulus,
(8) linear distance in Mpc, and (9) mean metallicity of the RGB with random 
and systematic errors divided by comma.

  Some additional comments about the galaxy properties are briefly
discussed below. The galaxies are listed in order by increasing
Right Ascencion.

{\bf ESO 215-09, KKs 40.} This is an isolated dIrr galaxy of low surface
brightness. It is noted in the Catalog of Neighboring Galaxies
(Karachentsev et al. 2004 = CNG) as an object of very high hydrogen
mass-to-luminosity ratio. Warren et al. (2004) performed surface
photometry of the galaxy in $B,V,R,I$- bands and derived its HI line
velocity field using ATCA. According to these data, the HI disk of
ESO~215-09 extends over 6 Holmberg radii, and the ratio of hydrogen
mass to $B$-luminosity is $(22\pm4) M_{\sun}/L_{\sun}$.



{\bf ESO 381-20.} HI mapping of this irregular galaxy was carried out
by C\^ot\'e et al. (2000). On our ACS frames the galaxy is resolved into
more than 20 000 red and blue stars. Given its $M_{HI}/L_B =
2.6 M_{\sun}/L_{\sun}$, ESO~381-20  has a considerable amount of gas
and hence a strong potential to form new stars.


{\bf KK 182, Cen 6.} This irregular galaxy of triangular shape was proposed 
as a member of the Cen~A/M~83 complex by
C\^ot\'e et al.(1997) but it turns out to be slightly to the background.

{\bf ESO 269-058.} This is a peculiar galaxy of I0 type with dusty patches.
NED gives an erroneous velocity $+$1853 km s$^{-1}$.
It is seen in its CM diagram (Fig. 4) that the vast majority
of $\sim$150 000 detected stars belong to the RGB population. The hydrogen
mass-to-luminosity ratio, $0.07 M_{\sun}/L_{\sun}$, turns out to be much lower
than typical values for irregular galaxies.

{\bf KK 189.} This dwarf spheroidal galaxy was not detected in the
HI line by Huchtmeier et al. 2001). Jerjen et al. (2000b) did
photometry of KK 189 in the $B,R$ bands, deriving the total color of
$B-R = 0.89$.

{\bf ESO 269-066, KK 190.} This is a dSph galaxy undetected in HI
(Huchtmeier et al. 2001). Its radial velocity, $+$784 km s$^{-1}$, was
measured via optical absorption lines by Jerjen et al. (2000a), who also
carried out surface photometry of ESO~269-66 in $B,R$ bands. The
CM diagram of the galaxy derived by us (Fig. 4) displays a pronounced
branch of red giant stars with a considerable metallicity scatter, 
unusual for galaxies of low luminosity ($-$13.6 mag).

{\bf KK 196, AM1318-444.} Surprisingly, this dIrr galaxy is still undetected
in HI. Its optical spectrum shows emission lines with the mean velocity
$+$741 km s$^{-1}$ (Jerjen et al. 2000b).

{\bf KK 197.} This dSph galaxy, undetected in HI, is situated 48$\arcmin$
away from NGC~5128 = Cen~A. Its CM diagram (Fig. 4) displays a large scatter
of colors for the stars near the tip of RGB. Our initial explanation of this
atypical feature assumed the presence of a number of high metallicity
stars projected from the periphery of Cen~A onto the KK~197 field.
However, the distribution of very red stars (situated to the right of the
tip of RGB) shows a clear concentration within the boundary of KK~197.
This implies that the observed scatter of colors (metallicities) of RGB
stars is a property of the dwarf galaxy itself and not
caused by outlying RGB stars in Cen~A.  Surface photometry
of KK~197 in $B,R$ bands was carried out by Jerjen et al. (2000b).

{\bf KKs 55.} This dSph galaxy is the nearest known companion to Cen~A
at a distance of 39$\arcmin$ in projection. CM diagram of KKs~55 exhibits
that the vast majority of the detected stars belong to the RGB with
small dispersion in their colors. The derived distance to KKs~55,
3.94 Mpc, coincides within errors with the distance to Cen~A, 3.77 Mpc.

{\bf IC 4247, ESO 444-34.} This is a dIrr galaxy of rather high surface
brightness. Judging from the derived distance, 4.97 Mpc, IC~4247 is
a companion to the bright spiral galaxy NGC~5236 = M~83.


{\bf NGC 5237.} The compact reddish galaxy of I0 type has an irregular
central part with an extended blue knot on the NW side
and a smooth periphery. The CM diagram of NGC~5237
displays about 100~000 stars, mainly belonging to the RGB. The blue stellar
population of the galaxy is concentrated towards its core.

{\bf KKs 57.} This galaxy of very low surface brightness is classified
as type dSph or dSph/Ir. It is not detected in the HI line by
Huchtmeier et al. (2001). KKs~57 is one of the faintest known
companions to Cen~A with $M_B = -10.3$ mag.


{\bf HIPASS1348-37.} This is a dIrr galaxy found in the blind HI survey
of the southern sky with the Parkes Telescope (Banks et al. 1999). The 
coordinates presented in NED are rather inaccurate.

{\bf ESO 383-87.} In our list, this bright ($B = 11.0$ mag) spiral galaxy of
SBdm type has the lowest radial velocity, $V_{LG} = +108$ km s$^{-1}$.
Amazingly, it was never resolved into stars before. Kemp \& Meaburn (1994)
noted that the galaxy is embedded in an extensive fairly spherical halo seen
down to the level of 2\% of the sky brightness. Unfortunately, ESO~383-87
was imaged with ACS in the F814W filter only, because of an HST guiding
problem. Another F814W image of the galaxy has been obtained with WFPC2
(GO \#8599). We carried out photometry of both the images and obtained
an $I$-band luminosity function of ESO~383-87. Applying
the edge-detection Sobel filter to the luminosity function, we derived the 
TRGB positions
23.76 (WFPC2) and 23.78 mag (ACS), which yields the galaxy distance to be
$3.45\pm0.34$ Mpc.  It is by far preferable to have the color information
provided by a $V$ filter observation but in this case the onset of the
red giant branch is sufficiently distinct that the distance measure based
on the $I$ observation alone is considered reliable.
Basing on the measured TRGB distance,
we assign the galaxy to be a companion to Cen~A.

{\bf HIPASS1351-47.}  This dIrr galaxy of low surface brightness was
found in the HIPASS survey (Banks et al. 1999). Its coordinates given
in NED differ from true ones by $2\farcm5$ . The large TRGB distance to
the galaxy together with its low radial velocity, $V_{LG} = 292$ km s$^{-1}$,
indicates that the galaxy is located behind the Cen~A/M~83 complex and
has a noticeable peculiar velocity towards the complex.

{\bf ESO 384-016.} This is a galaxy of dS0/Im type with a diffuse halo
around a compact central core. Jerjen et al. (2000a,b) performed
surface photometry of the galaxy in $B,R$ bands and estimated its
distance via surface brightness fluctuations to be 4.23 Mpc.  Our
estimate of the galaxy distance from the TRGB, 4.53 Mpc, is in good
agreement.

{\bf ESO 223-09.} This isolated dIrr galaxy is situated behind and East
of the Cen~A/M~83 complex in a zone of strong extinction. The galaxy has
a lot of gas ($2.2 M_{\sun}/L_{\sun}$) for continuing star formation.
It is seen in the CMD of Fig.~4 that this galaxy has a pronounced
intermediate age asymptotic giant branch population.  These stars lie 
immediately above the red giant branch but the onset of the RGB is 
sufficiently dominant that the TRGB can be clearly distinguished.

{\bf ESO 274-01, RFGC 2937.} This is an isolated Sd galaxy seen edge-on.
With its angular dimension of $13\farcm4\times1\farcm3$, only a small part of
the galaxy is situated within the ACS field of view. Nevertheless, our
photometry reveals more than 100,000 stars, the majority belonging
to the RGB. The measured TRGB distance, 3.09 Mpc, together with the galaxy
radial velocity, $V_{LG}=+335$ km s$^{-1}$, may indicate a peculiar motion of
ESO~274-01 away from us toward the Cen~A/M~83 complex.

{\bf ESO 137-18.} This is an isolated galaxy of type Sm or Im in the zone
of the Milky Way $(b = -7.\degr4$). In spite of significant contamination
by foreground stars, its CM diagram displays the RGB population,
yielding the galaxy distance of 6.40 Mpc.  As with ESO 223-09, there is a
substantial intermediate age population that gives rise to a well
populated asymptotic giant branch.  Nonetheless the TRGB is easily
identified.

  Apart from the galaxies discussed above, we also observed the galaxy
PGC 47885 reported to have a radial velocity of +570 km s$^{-1}$. 
This object turns
out to be a distant spiral galaxy unresolved into stars. According to
recent 6dF data (see NED), its radial velocity is +13848 km s$^{-1}$.

\section{The turn-over radius of the Cen~A/M~83 complex}

  The list of all known galaxies in a wide vicinity within a radius of
$\sim$4 Mpc around the centroid of the complex Cen~A/M~83 is presented in
Table 2. The list contains 87 galaxies. Some galaxies (for instance,
DDO 161) have radial velocities $V_{LG} < 550$ km~s$^{-1}$, but they have no
individual distance estimates. For other galaxies (for instance, the dwarf
spheroidal system KK~197) the distances are measured with high accuracy,
but the radial velocities are lacking. The columns of Table 2 contain the
following data: (1) galaxy name; (2) equatorial coordinates for the
epoch J2000; (3) morphological type; (4) ``tidal index'' following from
the Catalog of Neighboring Galaxies (Karachentsev et al 2004 :CNG):  
for every galaxy ``$i$'' we found its ``main disturber''(=MD), producing the
highest tidal action
$$\Theta_i = \max  \{\log(M_k/D_{ik}^3)\} + C,\;\;\;(i = 1, 2...  N)$$ 
where  $M_k$ is the total mass of any neighboring potential MD galaxy
(proportional to its luminosity with $M/L_B = 10 M_{\odot}/L_{\odot}$)
separated from the considered galaxy by a space distance $D_{ik}$;
the value of the constant $C$  is chosen so that
$\Theta=0$ when the Keplerian cyclic period of the galaxy with respect
to its MD equals the cosmic Hubble time, $1/H_{0}$; therefore 
positive values correspond to galaxies in groups, while the negative ones
correspond to isolated galaxies; (5) radial velocity of the galaxy
with respect to the Local Group centroid and its error; (6) distance to the
galaxy (in Mpc) and its error; (7) method of estimating the distance (``cep''
--- from cepheids, ``rgb'' --- from the tip of RGB, ``sbf'' --- from
surface brightness fluctuations, ``mem'' --- from  probable membership in
the known groups, ``h'' --- from radial velocity with the global Hubble
constant $H_0 = 72$ km s$^{-1}$Mpc$^{-1}$, Freedman et al. 2001);
(8) reference to the source of data on the distance or a new radial
velocity; and (9) notes regarding galaxy membership in the Cen A ("C")
and M 83 ("M") groups based on a positive tidal index with respect to
Cen A or M 83. 

The greater part of the data on galaxy distances
are taken from the CNG although more than 40\% of estimates
have appeared over the last two years. Apart from 24 new distance
measurements from Table 1 (=K06b), we have used distance estimates of
galaxies based on TRGB measurements by Tully et al. (2006)= T06, 
Karachentsev et al. (2006a)= K06a, Rejkuba (2004)=R04, {\ bf Sakai et al. (2004),}
Galazutdinova (2005) = G05, and Sharina (2005)= Sh05. New radial
velocities for 5 galaxies from the HIPASS survey were taken from
Koribalski et al. 2004 (=Ko04) and Meyer et al. 2004 (=M04). For two
dSph galaxies, KK 211 and KK 221, new optical radial velocities were
recently measured by Puzia \& Sharina (2006) via globular clusters. For the
spiral galaxy Circinus situated in the Zone of Avoidance, we have
determined its distance modulus from the Tully-Fisher relationship in
the $J,H$ and $K$ bands based on the data of the 2MASS survey. There is
another probable member of the Cen A group, ESO 270-17 = RFGC 2603.
This large ($15\arcmin$ diameter) SBm type galaxy seen edge-on has $V_{LG} =
583$ km s$^{-1}$, a bit exceeding our velocity cutoff. We have derived its
TF distance modulus based on B, R, I magnitudes to be $28.18\pm0.37$ that
comfortably puts ESO 270-17 in the Cen A suite.

Because of it's luminosity and probable dynamic importance, we give
special attention to the giant spiral galaxy NGC 5236 = M 83.
In the CNG it was listed with the cepheid distance $4.47\pm0.30$ Mpc (Thim et
al. 2003), which positions M 83 to the foreground of all its companions with
TRGB distances. We
ran DOLPHOT photometry on
archival ACS images of the M83 halo (program 9864),
The resulting CMD is shown in Fig.~5.  There is some uncertainty
because of the spread of stellar metallicity but the cepheid distance
seems underestimated.
Applying the edge-detection Sobel filter, we derived the TRGB
position to be I(TRGB) = $24.64\pm0.15$ within a color range of
$1.0 < (V - I) < 2.0$, which yields the galaxy distance $5.16\pm0.41$ Mpc.
On the northern edge of M 83 there is a faint elongated arc, KK 208, a
possible satellite dSph galaxy disrupted by M 83's tidal forces. Its
TRGB distance, $4.68\pm0.47$ Mpc, is at middle of the M83 cepheid and TRGB
distances.

  The overall distribution on the sky of the galaxies from Table~2 is
presented in Fig.~6. The galaxies with individual distance estimates are
shown by circles, while the galaxies with distances obtained from the
Hubble relation are marked by squares. To attach an impression
of depth to this distribution, the galaxy radial velocities
are given colors following the scale on the right side of the figure. 

  A 3-dimensional view of the Cen~A/M~83 complex and its surroundings is
presented in Fig.~7 where brighter galaxies are shown as larger balls.
Spiral -- irregular galaxies, and elliptical -- dwarf spheroidal galaxies
with accurate ("cep" and "rgb") distances are drawn as blue and red balls,
respectively.

  The Hubble diagram for galaxies in the Cen~A/M~83 complex and its closest
surrounding is shown in Fig.~8. Galaxies in groups and in the general
field are depicted by open and filled circles, respectively. Galaxies
for which Cen~A and M~83 are main attractors are joined with these
dominant systems by solid lines. Both the radial velocities and the
distances are given in relation to the Local Group centroid. The two
dashed lines correspond to alternate choices of the Hubble parameter, 
the global value of
72 km s$^{-1}$ Mpc$^{-1}$, (Freedman et al. 2001) and the mean local value
68 km s$^{-1}$ Mpc$^{-1}$, (Karachentsev et al. 2006),
curved because of the decelerating action of the
Local Group with a total mass of $1.3\cdot 10^{12} M_{\sun}$.

As we had come to anticipate,
field galaxies show a markedly lower peculiar
velocity dispersion than that of the members of the groups. Field galaxies
are also seen to have a tendency to move toward the Cen~A/M~83 complex.
This motion is manifested as the known wave effect around an attractor:
galaxies on the near side of the complex have velocities higher than
the Hubble expectation and galaxies on the far side have velocities 
lower than the Hubble expectation. The distances and velocities of the
main galaxies, Cen~A and M~83, coincide within errors with the mean
distance and the mean velocity of their companions 
That is, Cen~A and M~83 are not only the centers of both groups
according to their position in the sky, but also the dynamical centers
of these groups. Here, the peculiar velocities of the group centroids
with respect to the mean local Hubble flow at
$H_0 = 68$ km s$^{-1}$ Mpc$^{-1}$ do not exceed 25 km s$^{-1}$.

  A rough estimate of the radius of the zero-velocity sphere, $R_0$, can
be made considering the mutual motion of the main members of the groups,
Cen~A and M~83. The spatial
separation between them is $R_{gg} = (1.73\pm0.56$) Mpc.
Assuming that their mutual tangential motion is zero,
the difference of their radial velocities projected onto the line joining
the galaxies is equal to $(V_{M83} - V_{CenA}) = +35\pm6$ km~s$^{-1}$.
It would take a tangential motion of
69~km~s$^{-1}$ of the two galaxies toward each other for one to see
the other at zero velocity.
If the galaxies M~83 and Cen~A are receding from each other, as is 
suggested but not certain,
then $R_0$ is less than $R_{gg}$. 
Alternatively, consider the motions
of the group centroids. We obtained from Table~2 the mean values $\langle V \rangle =
292\pm37$ km s$^{-1}$ and $\langle D \rangle = 3.76\pm0.05$ Mpc for the Cen~A group, and
$\langle V \rangle = 318\pm22$ km s$^{-1}$ and $\langle D \rangle = 4.79\pm0.10$ Mpc for the M~83
group. Under the same assumption that the relative tangential motions of the 
groups
are zero, we obtain that the centers of groups are moving apart from one 
another
at a velocity of $+70\pm30$ km s$^{-1}$ at a mutual spatial separation
$R_{cc} = (1.43\pm0.11)$ Mpc.  

  A more detailed approach to determining the radius $R_0$ was used by
Karachentsev \& Kashibadze (2006). They considered the velocity field
around the Local Group
of galaxies with accurate distance estimates
and obtained the value $R_0(LG) = 0.96\pm0.03$ Mpc. By varying the
mass center between the two main members of the LG and seeking a
minimum scatter of galaxies on the Hubble diagram, Karachentsev \&
Kashibadze (2006) estimated a mass ratio of 0.8:1.0 for our Galaxy
and M~31, in accordance with the observed ratio of luminosities of these
galaxies. A similar analysis of the velocity field of the galaxies around
the M~81 group yielded the value of $R_0(M81) = 0.89\pm0.05$~Mpc
and the ratio of masses of two main galaxies, M~81 and M~82, of
1.0:0.5, in good agreement with the ratio of the luminosities
of these galaxies.

  Examining the Hubble pattern around the complex Cen~A/M~83, we suppose
the center of mass of the complex to be coincident with Cen~A,
then we determine velocities $V$ and distances $R$ of the galaxies given in
Table~2 with respect to Cen~A. Here, we take into account only the objects
whose separations from Cen~A along the line of sight exceed their
projected separations on the sky: $|D_g - D_c| > R_p$, and also excluded
close companions ($ R < 1$ Mpc) in order to reduce the contribution
from virial motions. At total, there are 18 galaxies with spatial distances
from Cen~A, $1 < R < 4$ Mpc which are comfortably situated along a line
of sight passing through Cen~A. Surprisingly, all of them
reside behind the Cen A group.
The resulting Hubble diagram for them is presented in Fig.~9. Each galaxy
is shown by a circle with horizontal and vertical bars denoting
standard errors of distance and velocity. Galaxies in groups
( $\Theta > 0$) and in the general field ($ \Theta < 0$) are indicated by
open and filled circles, respectively. The solid line in Fig.~9 corresponds
to the Hubble regression with $H_0 = 72$ km s$^{-1}$ Mpc$^{-1}$ and
the $R_0$ value which ensures the minimum scatter of galaxies with respect
to the Hubble regression. As is seen, the regression line crosses the
zero-velocity line at $R_0 = 1.40\pm0.11$ Mpc, where the standard
deviation, 0.11 Mpc, is a bootstrap estimate. The mean-square peculiar
velocity of the galaxies relative to the homogeneous Hubble flow is
32 km s$^{-1}$. However, this value is affected to a considerable
degree by measurement errors in galaxy distances. After quadratic
subtraction of observational errors, the typical peculiar velocity drops
to 2 km s$^{-1}$.

  Thus, the Hubble flow around the Cen~A/M~83 complex proves to be rather
cold. Low velocities of chaotic motions $\sim(10 - 20)$ km s$^{-1}$ are
characteristic also of the surroundings of two other nearby complexes:
the Local Group and the M~81 group. This feature is direct evidence of
the existence of dark energy which dominates at distances $\ga 1$ Mpc
from the center of a typical loose group of galaxies.

  \section{ Mass estimates of the complex Cen~A/M~83}

The separation of galaxies in the Centaurus region into
the Cen~A and M~83 groups was
considered in detail by Karachentsev et al. (2002b).  They measured masses
for the two groups from the virial theorem  $$M_{vir} =
 3\pi N \times(N-1)^{-1} \times G^{-1} \times \sigma^2_v \times R_H,$$
where $\sigma_v^2$ is the dispersion of radial velocities with respect to
the group centroid, and $R_H$ is the mean projected harmonic radius, or from
the orbital motions of companions around the principal galaxy,
 $$M_{orb} =
(32/3\pi)\times G^{-1}\times(1- 2e^2/3)^{-1} \langle R_p\times \Delta V^2_r\rangle,$$
assuming arbitrarily oriented Keplerian orbits of companions with the
mean eccentricity of galaxy orbits postulated to be $e$ = 0.7.  New
observational data modifies the mass estimates. Basing on
radial velocities and mutual separations of supposed members of Cen~A
group (marked in the last column of Table~2 as ``C''), we obtained the
mean harmonic radius of the group $R_H = 192 $~kpc and the radial
velocity dispersion $\sigma_v = 136 $ km s$^{-1}$, which yields a virial
mass of the group $M_{vir} = 8.1 \cdot 10^{12} M_{\sun}$.
The orbital mass estimate turns out to be somewhat smaller,
$M_{orb} = 6.4 \cdot 10^{12} M_{\sun}$.

  The new values of the harmonic radius and radial velocity dispersion for
the M~83 group are 89~kpc and 61 km s$^{-1}$, respectively. Mass estimates
for the M~83 group, $M_{vir} = 0.82\cdot 10^{12} M_{\sun}$ and
$M_{orb} = 0.89\cdot 10^{12} M_{\sun}$, are almost an order of magnitude
lower than for the Cen~A group. 

The total blue luminosities
of these groups, $L_B$(CenA) = $6.0\cdot 10^{10} L_{\sun}$ and
$L_B$(M~83) = $2.5\cdot 10^{10} L_{\sun}$, differ from one another
much less than the mass estimates. This situation is consistent
with the proposition (Bahcall et al. 2000; Tully 2005) that giant
elliptical galaxies, as well as groups with a predominantly elliptical
population, have masses per unit luminosity about 3 times higher than giant
spirals or groups with predominantly spiral populations.
In the present case, the elliptical dominated Cen~A group has a mass-to-light 
ratio $M/L_B = 125 M_{\sun}/L_{\sun}$ while the spiral dominated M~83 group
has $M/L_B = 34 M_{\sun}/L_{\sun}$.
This difference cannot be attributed only to the presence in spiral galaxies 
of a young (blue) stellar population and dust clouds since the infrared
luminosities in the $K$- band, $L_K$(CenA) =$15.1\cdot 10^{10} L_{\sun}$ and
$L_K$(M~83) =$7.1\cdot 10^{10} L_{\sun}$, taken from 2MASS, also
differ far less than the mass estimates of the groups.

Consideration of
galaxy motions in the vicinity of the Cen~A/M83 complex gives us a mass
estimate on a scale of  $\sim$1 Mpc.
According to Lynden-Bell (1981) and
Sandage (1986), the total mass of a group is expressed via the turn-over
radius $R_0$ and the age of the universe $T_0$ as
\begin{equation}
   M_T = (\pi^2/8G)\cdot R_0^3 \cdot {T}_0^{-2}.
\end{equation}
where $G$ is the gravitation constant. 
In the ``concordant'' flat cosmological model with a non-zero
$\Lambda$- term, equation (1) needs to be modified as
\begin{equation}
   M_T = (\pi^2/8G)\cdot R_0^3 \cdot H_0^2 \cdot f(\Omega_m)^{-2},
\end{equation}
where
\begin{equation}
f(\Omega_m) = 1/(1-\Omega_m) -
	    0.5\cdot\Omega_m\cdot (1-\Omega_m)^{-1.5} \cdot {\rm arccosh}[(2/\Omega_m) - 1]
\end{equation}
For $\Omega_m = 0.27$, we have $f(0.27) = 0.82$, and with $H_0 =
72$ km s$^{-1}$ Mpc$^{-1}$ corresponding to $T_0 = 13.7$ Gyr,
we obtain a new expression for the total mass:
  $(M_T/M_{\sun}) = 2.2\cdot 10^{12} (R_0/$Mpc$)^3$,
yielding $M_T = (6.0\pm1.4)\cdot 10^{12} M_{\sun}$.
Thus, the mass estimates of the Cen~A group
made from 
galaxy motions internal and external  to the group are in good agreement at
$6-8 \cdot 10^{12} M_{\sun}$.

  The outer halo of Cen~A has been recently studied by Peng et al. (2004)
based on the kinematics of 148 planetary nebulae situated at radii beyond
20 kpc. Applying different dynamical models, they obtained the total mass
of the galaxy within 80 kpc to be $(5.0 - 5.9)\cdot 10^{11} M_{\sun}$.
This mass is much lower than our estimates presented above. 
Part of the difference is due to the factor 5 difference in scale.
Another factor of 2--3 remains unexplained.  By contrast, we have good 
agreement with the dynamical study of globular clusters around Cen~A
by Woodley (2006). The motions of 340 globular clusters within a radius 
of 45~kpc imply a mass of $(0.8 - 1.8)\cdot 10^{12} M_{\sun}$.  We find
a mass 6 times greater on a scale 9 times greater.

We can make a comparison with a statistical measure of group masses.
  Over the last years, a new possibility of determining the total mass of
groups has appeared based on measuring a signal of weak lensing of more
distant galaxies. Hoekstra et al. (2005) detected such a signal by measuring
the orientation of the major axes of background galaxies around single
``lens galaxies'' with photometric redshifts $z = 0.2 - 0.4$.
The giant galaxies Cen~A and M~83 would look at these distances like
ordinary field galaxies. From the data by Hoekstra et al. (2005), galaxies
like Cen~A and M~83 with their blue luminosities 3.1 and
$2.3\cdot 10^{10} L_{\sun}$ generate a signal of lensing on a scale of
$\sim$0.5 Mpc which corresponds to a total mass of
$(3.8\pm1.4) \cdot 10^{12} M_{\sun}$. This estimate of mass is quite
consistent with our estimate of the total mass of Cen~A/M~83 made from the
distortion to the Hubble velocity field around the Cen~A/M~83 complex.

  \section{Concluding remarks}

  We have presented new distances to 24 galaxies situated in the nearby
binary group Cen~A/M~83 and its vicinity. A total of 87 galaxies are
presently known inside a sphere of radius $\sim$4 Mpc around the
centroid of the complex. Among them, 17 dwarf spheroidal galaxies
do not yet have measured radial velocities. Among 66 galaxies with
individual distance estimates, for 2 galaxies the distances
were measured via the luminosity of cepheids, for 61 galaxies the
distances were determined by the TRGB method, and for 3 galaxies the
distances were estimated from the Tully-Fisher relation or surface
brightness fluctuations. Thus, another 21 galaxies in the volume under
discussion are targets for future distance measurements.

  Considering galaxies only in close proximity to Cen~A and M~83, we
determined the total (virial or orbital) mass of the whole Cen~A/M~83 complex
to be $ (6.4 - 8.1) \cdot 10^{12} M_{\sun}$ with 90\% of the mass associated
with the component around Cen~A. An independent
estimate of the mass of the Cen~A component was made from the
distortion of the cosmic expansion velocity field among the galaxies 
surrounding
the complex. The observed deceleration of neighboring galaxies
caused by the mass of the Cen~A group is characterized by the
sphere of radius $R_0$ which separates the group from the general
cosmological expansion. The radius $R_0$ about Cen~A
lies in the range 1.29 -- 1.51 Mpc, which corresponds to a mass
$(6.0\pm1.4) \cdot 10^{12} M_{\sun}$ in a model with
$\Omega_{\Lambda} = 0.73$, $\Omega_m = 0.27$.  This mass estimate is
in excellent agreement with the mass estimates from the internal motions
in the Cen~A group.

Estimates of the total mass of Cen~A are listed in Table 3.
Ordered by scale length, they probe the scale range from 80 to
1400 kpc. Apart from the first mass estimate (PNe dynamics), the
remaining ones agree with each other within 2-$\sigma$ significance levels.
This agreement is satisfactory because every method is subject to
substantial uncertainties.  

The key result of our observations is 
the determination of $M/L_B = 125~M_{\sun}/L_{\sun}$ for the Cen~A group.
In the cases of nearby groups dominated by the giant spirals M31, M81, 
and M83,
the identical methods of calculating masses results in estimates of 
$M/L_B$ of 16, 32, and 34, respectively.  This evidence supports the
proposition that environments that have dynamically evolved to the 
end--state of elliptical galaxies have dark matter halos that manifest
less light than the less evolved environments of spirals.


 \acknowledgements{
Support associated with HST programs 9771 and 10235 was provided by NASA
through a grant from the
Space Telescope Science Institute, which is operated by the Association of
Universities for Research in Astronomy, Inc., under NASA contract
NAS5--26555. This work was also supported by RFFI grant 04--02--16115.
This study has made use of the NASA/IPAC Extragalactic Database (NED),
the HI Parkes All Sky Survey (HIPASS), and the Two Micron All-Sky Survey.}

{}

\newpage
\begin{figure}
\centerline{\psfig{figure=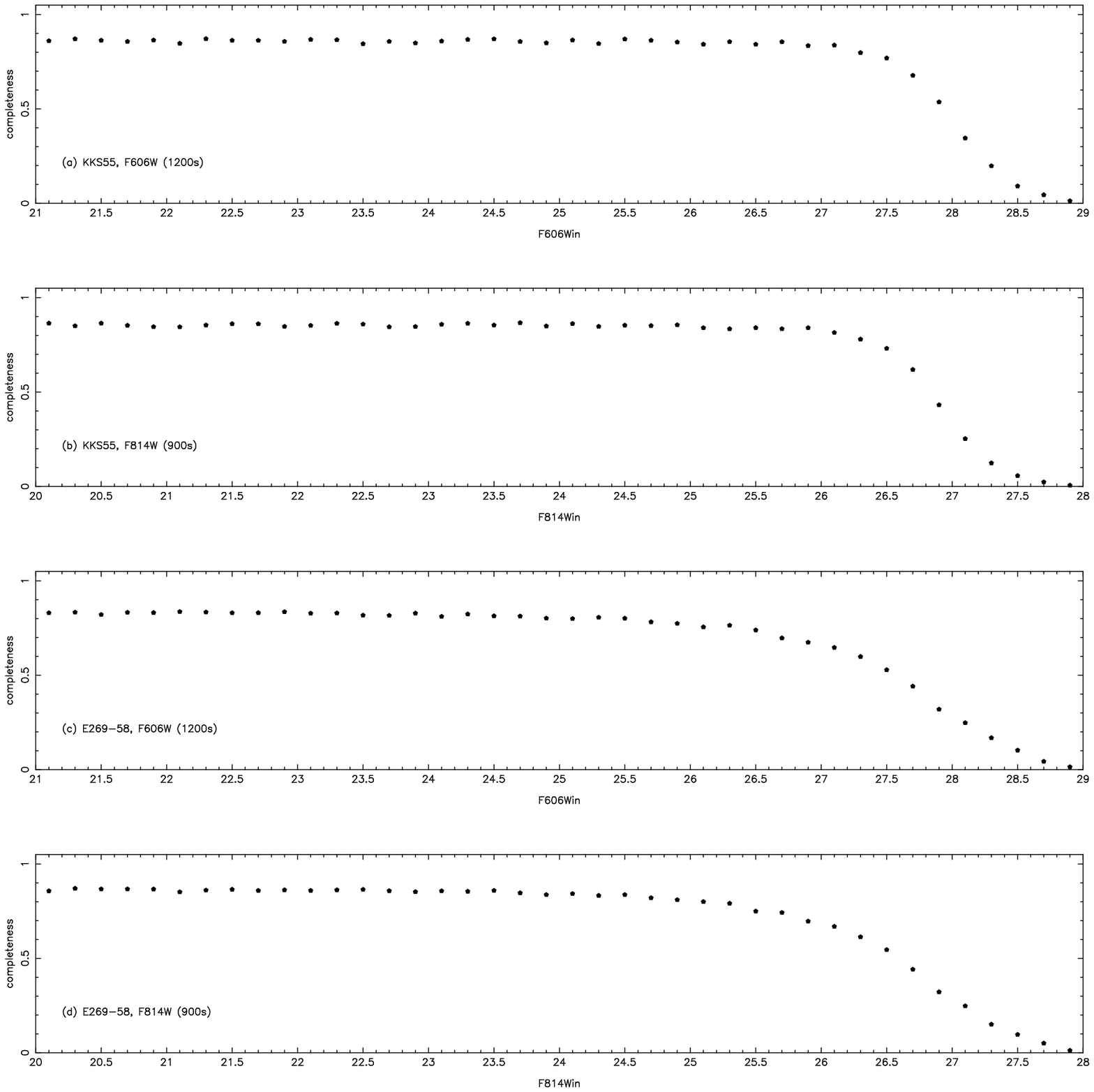,width=\textwidth,angle=0}}
\caption{Photometric completeness as a function of magnitude for the 
            least crowded
	    (KKs~55) and the most crowded (ESO~269--58) galaxies.}
\end{figure}

\begin{figure}
\centerline{\psfig{figure=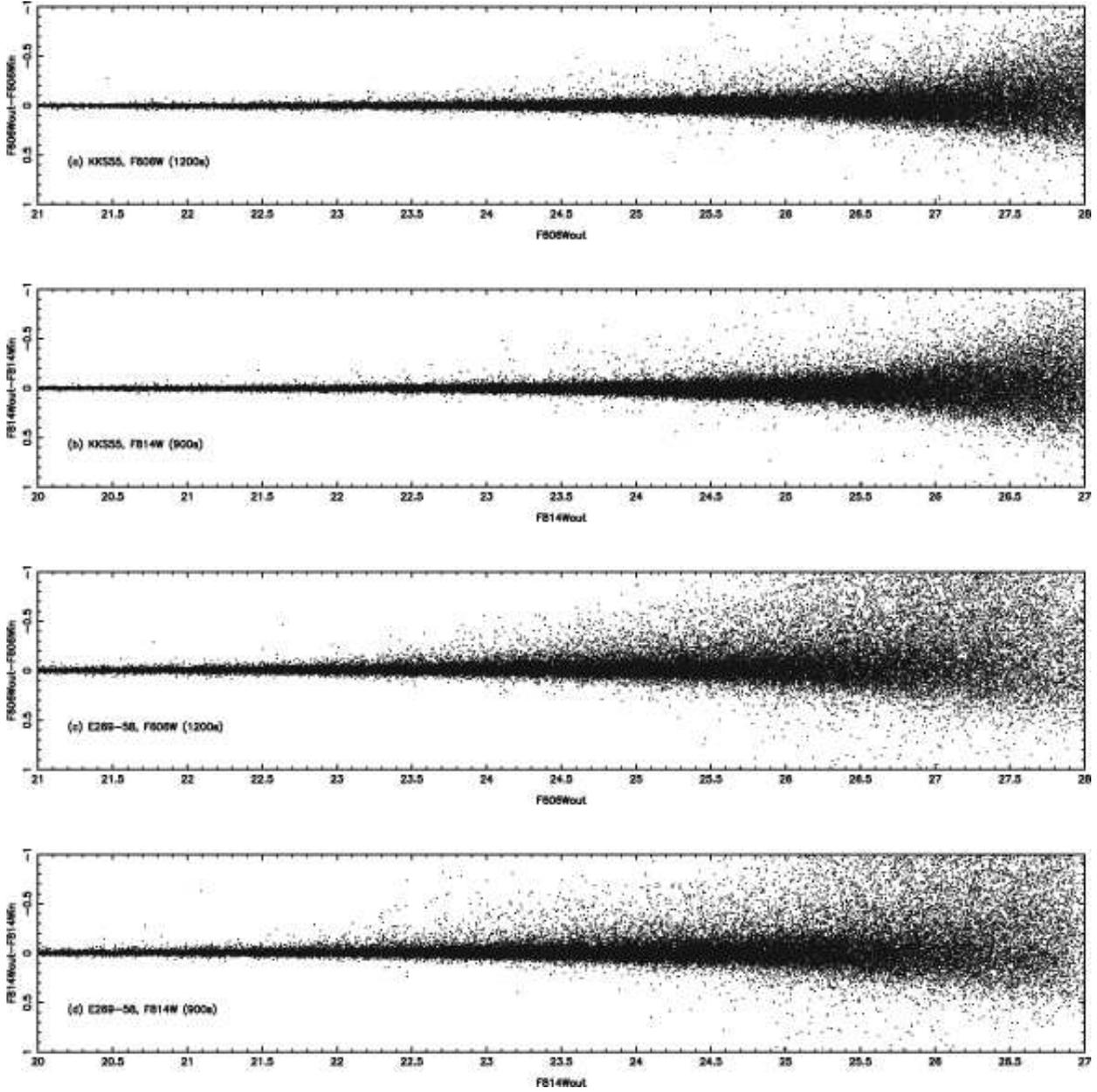,width=\textwidth,angle=0}}
\caption{Mean photometric error as a function of magnitude for KKs~55 
            and ESO~269--58.}
\end{figure}

\begin{figure}
\centerline{
\begin{tabular}{c}
\psfig{figure=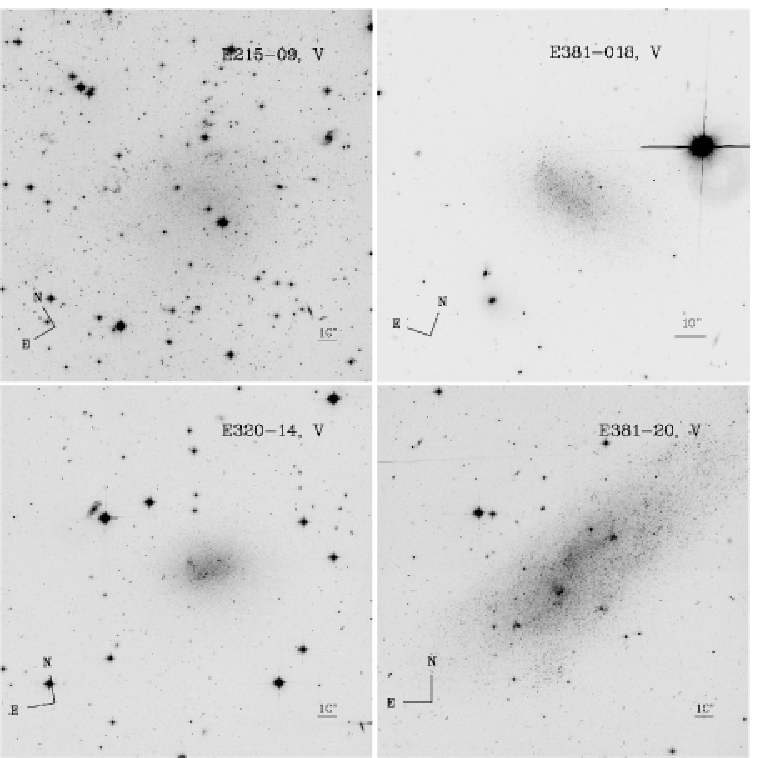,width=0.44\textwidth,angle=0} 
\psfig{figure=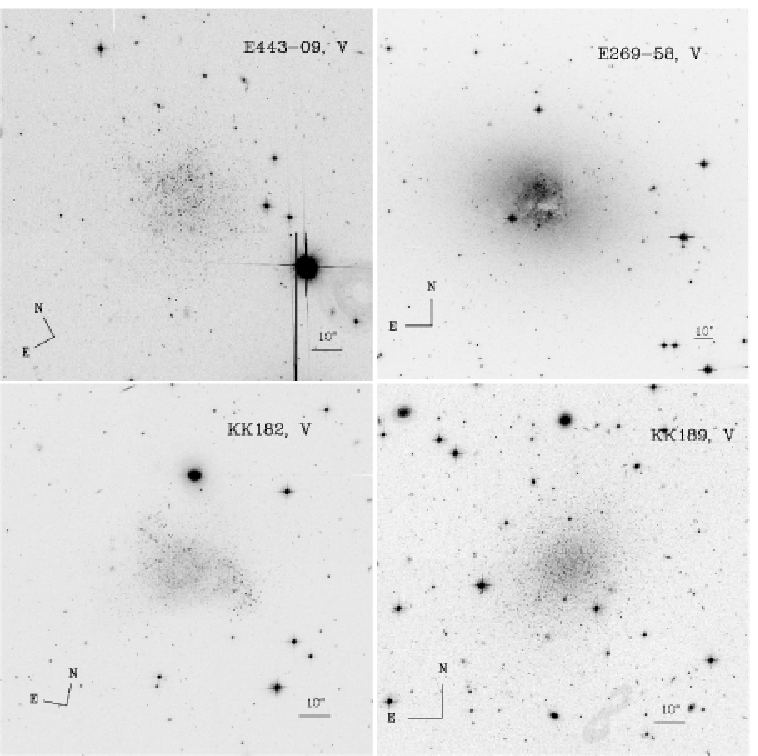,width=0.44\textwidth,angle=0}
\end{tabular}
}
\centerline{
\begin{tabular}{c}
\psfig{figure=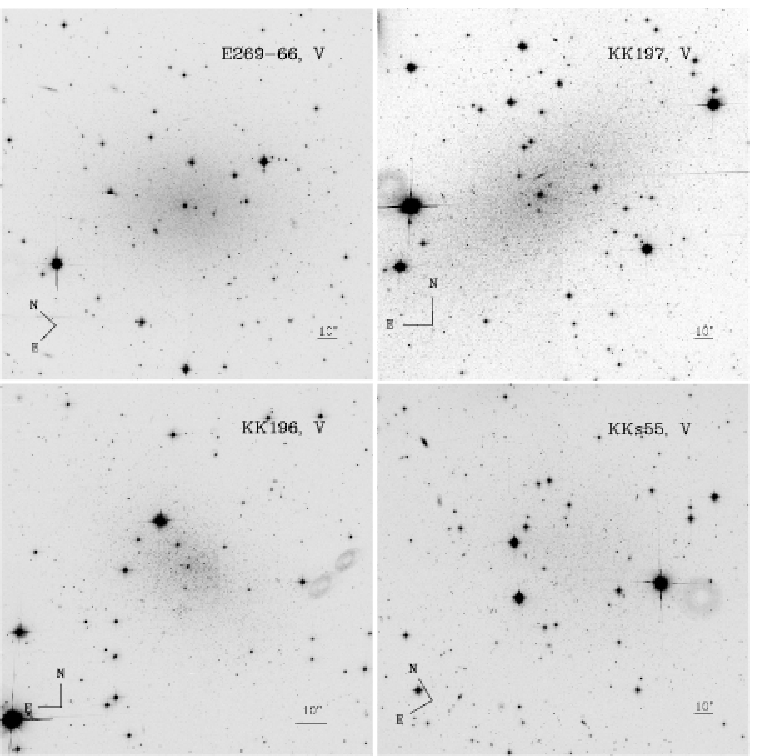,width=0.44\textwidth,angle=0}
\psfig{figure=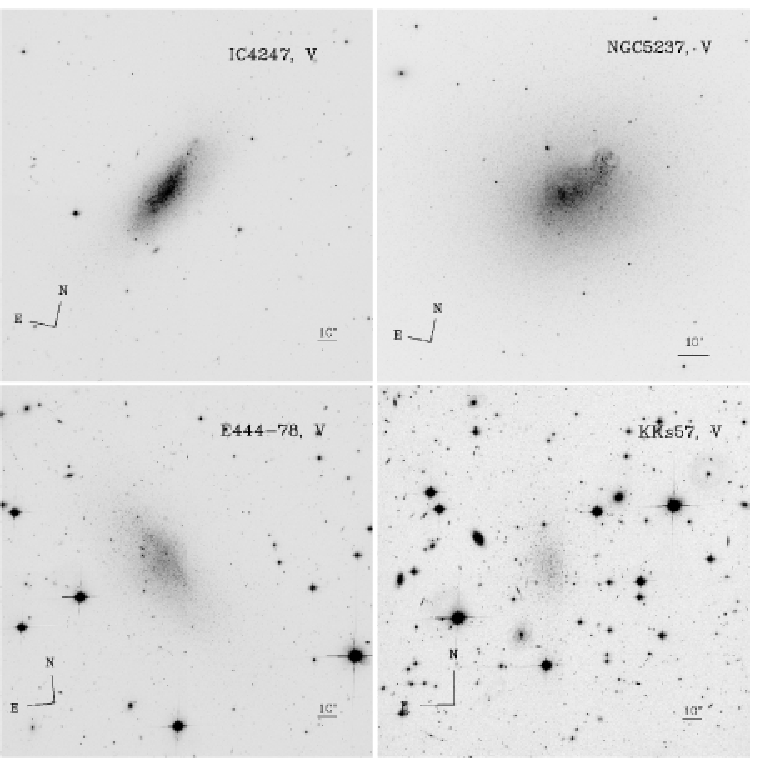,width=0.44\textwidth,angle=0}
\end{tabular}
}
\centerline{
\begin{tabular}{c}
\psfig{figure=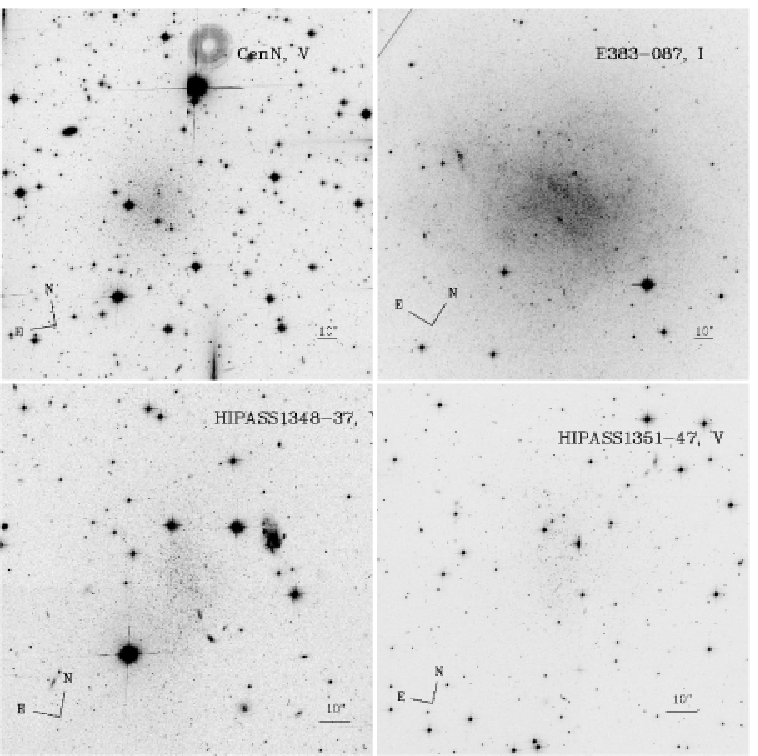,width=0.44\textwidth,angle=0}
\psfig{figure=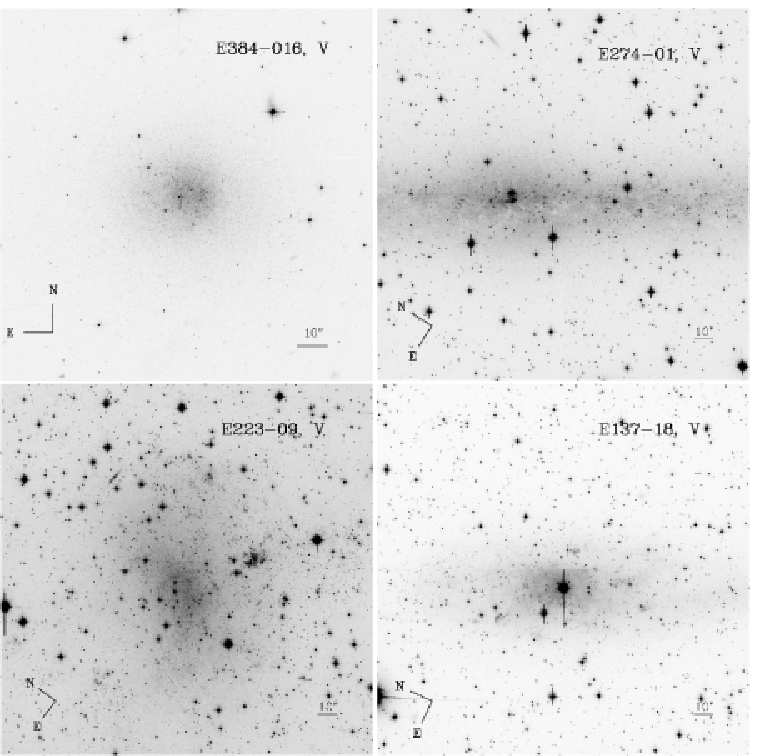,width=0.44\textwidth,angle=0}
\end{tabular}
}
\caption{ACS ``$V$'' images of 24 nearby galaxies; 23 from two 600 second
	    exposures in F606W and one (ESO 383-87) from two 450 second
            exposures in F814W.}
\end{figure}

\begin{figure}
\centerline{
\begin{tabular}{c}
\psfig{figure=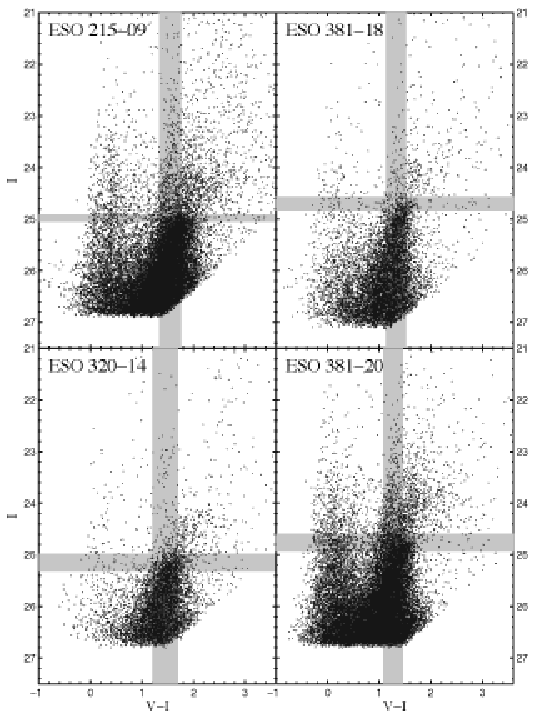,width=0.44\textwidth,angle=0} 
\psfig{figure=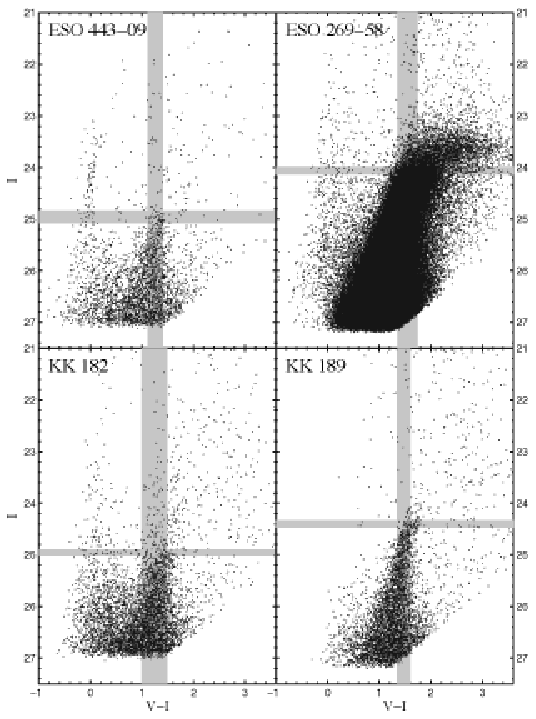,width=0.44\textwidth,angle=0}
\end{tabular}
}
\centerline{
\begin{tabular}{c}
\psfig{figure=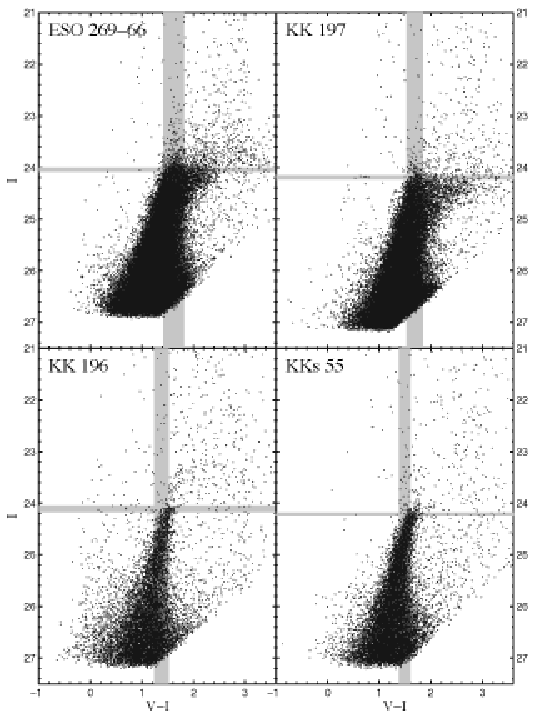,width=0.44\textwidth,angle=0}
\psfig{figure=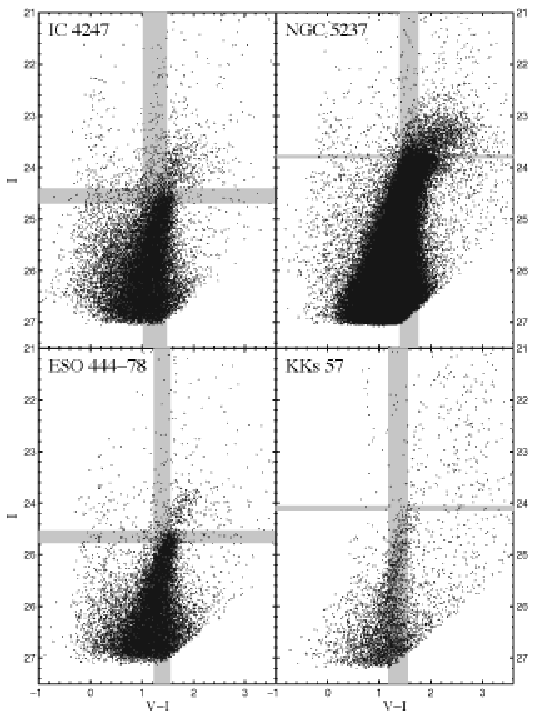,width=0.44\textwidth,angle=0}
\end{tabular}
}
\caption{ACS color-magnitude diagrams for 23 nearby galaxies in the
	    Cen~A/M~83 area. The uncertainties in the TRGB measurement and 
	    in the RGB mean color
	    $<V-I>_{-3.5}$ are shown as shaded regions.}
\end{figure}





\begin{figure}
\figurenum{4}
\centerline{
\begin{tabular}{c}
\psfig{figure=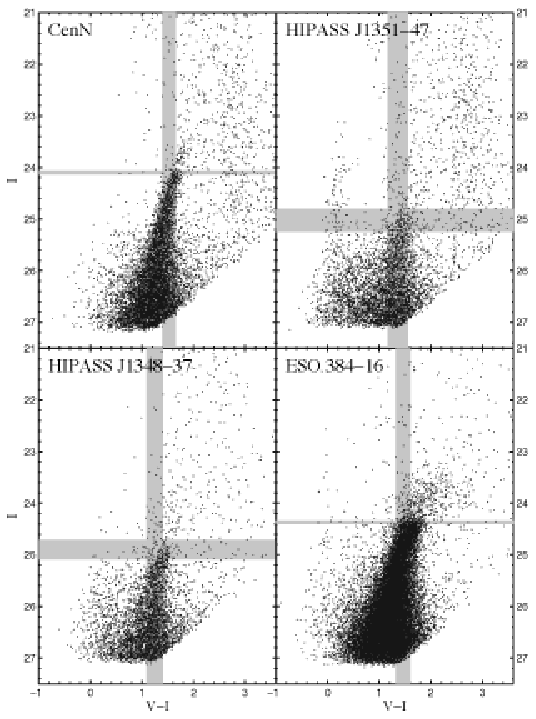,width=0.44\textwidth,angle=0} 
\psfig{figure=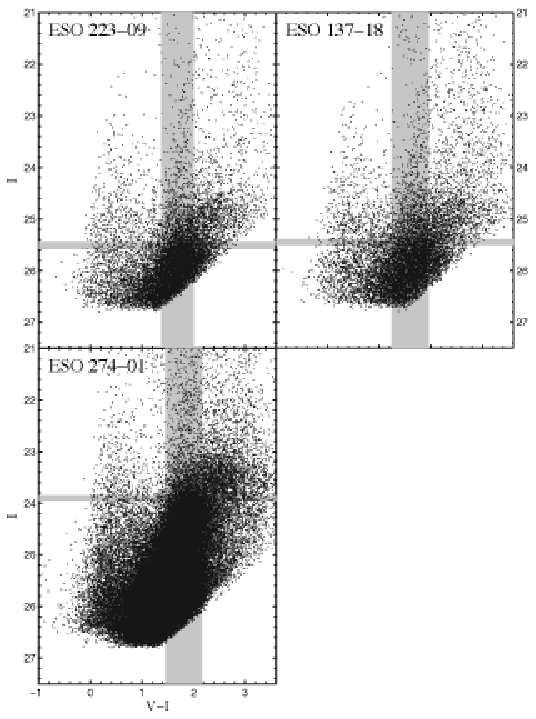,width=0.44\textwidth,angle=0}
\end{tabular}
}
\caption{continued}
\end{figure}



\begin{figure}
\figurenum{5}
\epsscale{0.8}
\psfig{figure=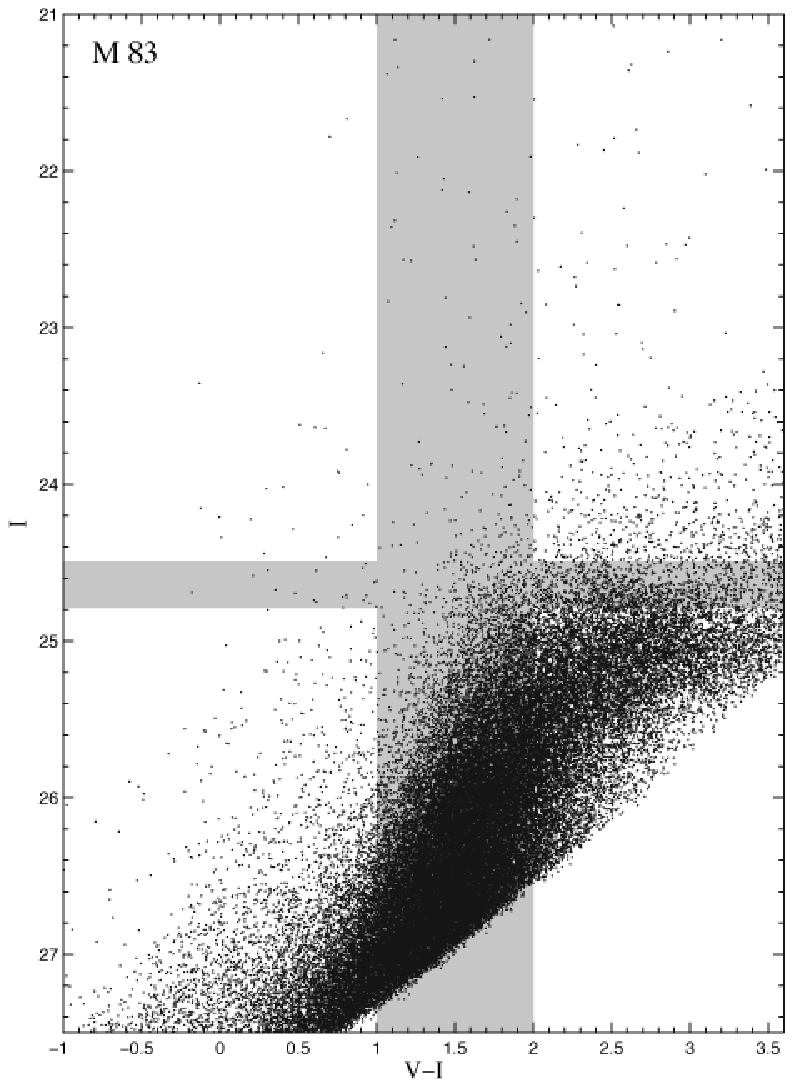,width=0.30\textwidth,angle=0}
\caption{ACS color-magnitude diagram for M83. The color measurement range
and TRGB magnitude uncertainty range are shown as shaded regions.}
 \end{figure}

\begin{figure}
\figurenum{6}
\epsscale{1.0}
\psfig{figure=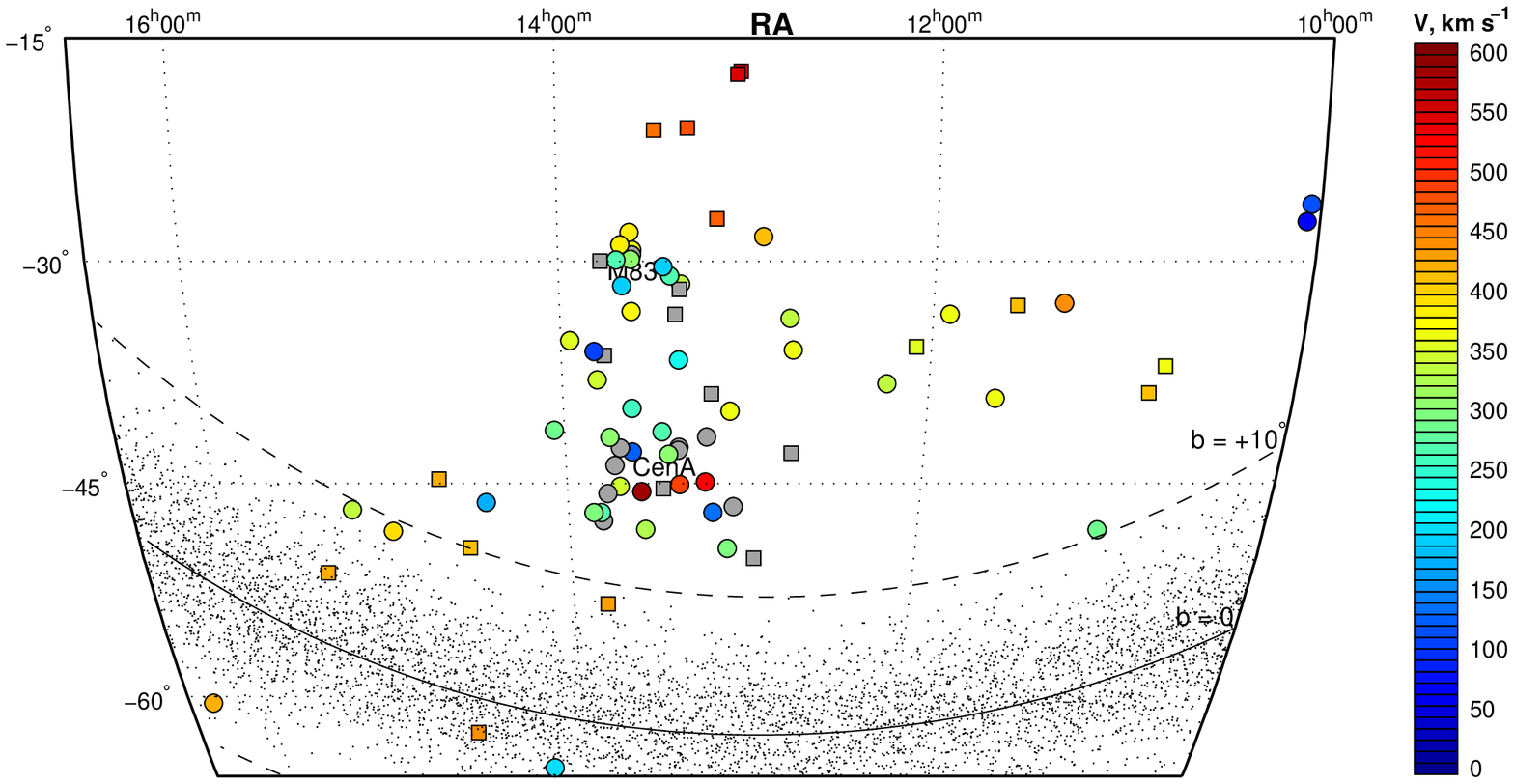, width=0.8\textwidth,angle=270}
\caption{The distribution of galaxies in and around the Cen~A/M~83 complex.
	    The galaxies with individual distance estimates are shown by
	    circles, while the galaxies with distances obtained from the
	    Hubble relation are marked by squares. Radial velocities of
	    galaxies are indicated by different colors. The Milky
	    Way zone of avoidance is shown as the dotted area.}
\end{figure}
\begin{figure}
\figurenum{7}
\epsscale{1.0}
\plotone{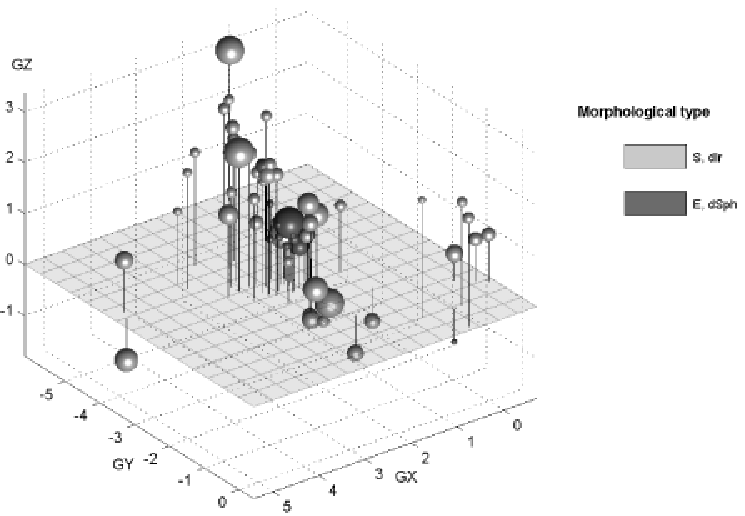}
\caption{The 3D- view of the Cen~A/M~83 complex. Galaxies of
	    young and old stellar population are shown by light and dark
	    grey, respectively.}

\end{figure}

\begin{figure}
\figurenum{8}
\epsscale{1.0}
\plotone{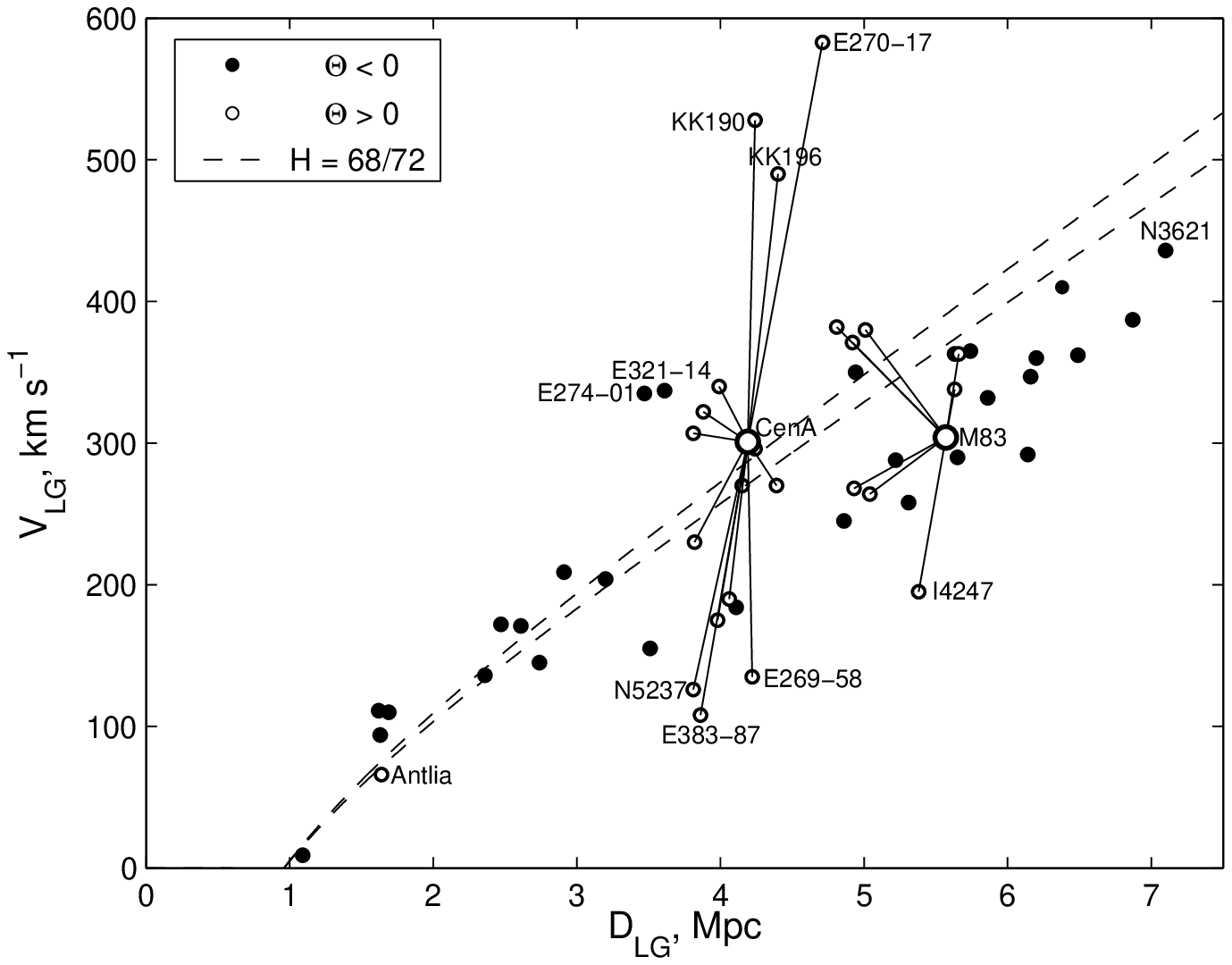}
\caption{The Hubble diagram for galaxies in the Cen~A/M~83 complex
	    and its vicinity. Galaxies in groups ( $\Theta > 0$) and in the
	    general field ($ \Theta < 0$) are indicated by open and filled
	    circles, respectively. The companions to Cen~A and M~83 are
	    connected by lines to the main galaxies. The two
	    dashed lines correspond to Hubble relations at $H_0 = 68$
	    and 72 km s$^{-1}$ Mpc$^{-1}$.}
\end{figure}

\begin{figure}
\figurenum{9}
\epsscale{1.0}
\plotone{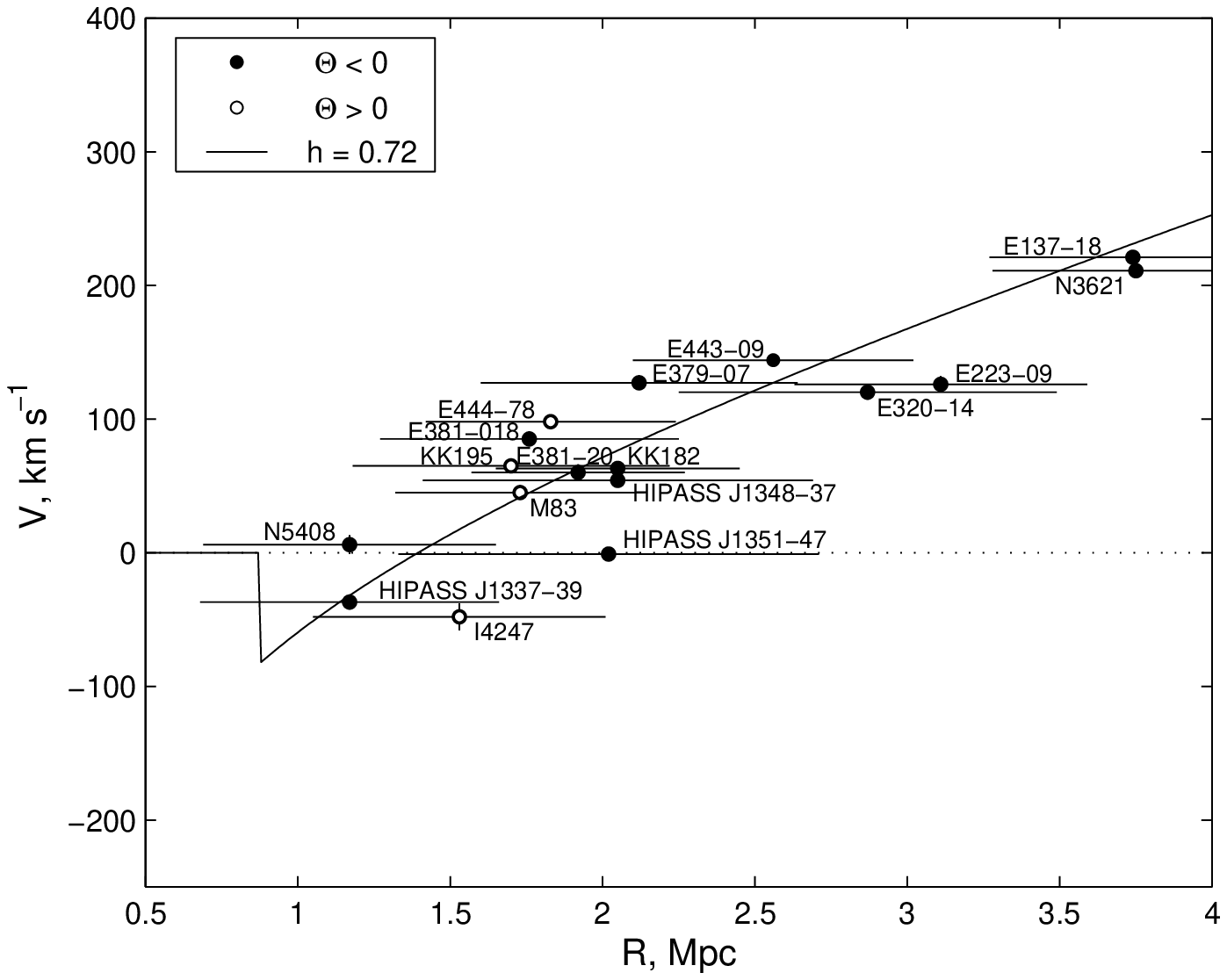}
\caption{Distribution of the radial velocity difference and the spatial
	    distance of galaxies with respect to the Centaurus A.
	    Galaxies in groups and in the general field are indicated by
	    open and filled circles, respectively.} 
\end{figure}

\hoffset=-1cm
\begin{table}
\caption{ New distances to galaxies in the Cen A/M 83 complex}
\medskip
\begin{tabular}{lcccccccc}
\hline
\hline
Name & RA(J2000.0)Dec & $V_{LG}$ & $I_{TRGB}$ & (V-I)$_{-3.5}$ & $A_I$ & $\mu_0$  & D & [Fe/H] \\
     &               &          & $\sigma_{TRGB}$ & $\sigma_{V-I}$ &  & $\sigma_{\mu_0}$ & & $\sigma_{[Fe/H]}$ \\  
\hline
E215--09,KKs40 & 105730.2$-$481044 &  290 & 24.98 & 1.55$\pm$0.006 & 0.43 & 28.60 & 5.25 & $-1.861$ \\
	       &                   &      &  0.07 & 0.21 &      &  0.16 & &0.024, 0.150     \\
E320--14,KKs44 & 113753.4$-$391314 &  362 & 25.15 & 1.45$\pm$0.012 & 0.28 & 28.92 & 6.08 & $-1.923$ \\
	       &                   &      &  0.17 & 0.25 &      &  0.22 &  &0.042, 0.154    \\
E381--18       & 124442.7$-$355800 &  353 & 24.70 & 1.32$\pm$0.011 & 0.12 & 28.63 & 5.32 & $-2.059$ \\
	       &                   &      &  0.14 & 0.20 &      &  0.20 &  &0.044, 0.162   \\
E381--20       & 124600.4$-$335017 &  332 & 24.76 & 1.27$\pm$0.007 & 0.13 & 28.68 & 5.44 & $-2.343$ \\
	       &                   &      &  0.02 & 0.19 &      &  0.14 &  &0.033, 0.177    \\
E443--09,KK170 & 125453.6$-$282027 &  410 & 24.96 & 1.26$\pm$0.018 & 0.13 & 28.88 & 5.97 & $-2.435$ \\
	       &                   &      &  0.12 & 0.15 &      &  0.17 & &0.083, 0.181    \\
	       &                   &      &       &      &      &       &   &  \\
KK182,Cen6     & 130502.9$-$400458 &  360 & 24.96 & 1.24$\pm$0.018 & 0.20 & 28.81 & 5.78 & $-2.741$ \\
	       &                   &      & 0.06  & 0.25 &      &  0.15 & &0.085, 0.196  \\
E269--058      & 131032.9$-$465927 &  142 & 24.06 & 1.55$\pm$0.003 & 0.21 & 27.90 & 3.80 & $-1.431$ \\
	       &                   &      & 0.08  & 0.20 &      &  0.16 & &0.007, 0.122  \\
KK189          & 131245.0$-$414955 &  $-$ & 24.40 & 1.47$\pm$0.012 & 0.22 & 28.23 & 4.42 & $-1.739$ \\
	       &                   &      & 0.07  & 0.13 &      &  0.16 & &0.042, 0.143  \\
E269--66,KK190 & 131309.2$-$445324 &  528 & 24.04 & 1.62$\pm$0.005 & 0.18 & 27.91 & 3.82 & $-1.220$ \\
	       &                   &      & 0.04  & 0.20 &      &  0.15 & &0.014, 0.105   \\
KK196          & 132147.1$-$450348 &  490 & 24.11 & 1.38$\pm$0.010 & 0.16 & 28.00 & 3.98 & $-1.961$ \\
	       &                   &      & 0.06  & 0.14 &      &  0.15 &  &0.034, 0.156  \\
	       &                   &      &       &      &      &       &   &  \\
KK197          & 132201.8$-$423208 & $-$  & 24.19 & 1.69$\pm$0.006 & 0.30 & 27.94 & 3.87 & $-1.190$ \\
	       &                   &      & 0.04  & 0.16 &      &  0.15 &  &0.017, 0.102  \\
KKs55          & 132212.4$-$424351 & $-$  & 24.21 & 1.49$\pm$0.006 & 0.28 & 27.98 & 3.94 & $-1.813$ \\
	       &                   &      & 0.03  & 0.11 &      &  0.14 &  &0.036, 0.147  \\
I4247,E444--34 & 132644.4$-$302145 &  195 & 24.55 & 1.25$\pm$0.008 & 0.12 & 28.48 & 4.97 & $-2.367$ \\
	       &                   &      & 0.15  & 0.23 &      &  0.21 &  &0.034, 0.178  \\
E444--78,UA365 & 133630.8$-$291411 &  363 & 24.65 & 1.39$\pm$0.005 & 0.10 & 28.60 & 5.25 & $-1.746$ \\
	       &                   &      & 0.10  & 0.16 &      &  0.17 &  &0.023, 0.143  \\
N5237          & 133738.9$-$425051 &  131 & 23.79 & 1.58$\pm$0.003 & 0.18 & 27.66 & 3.40 & $-1.330$ \\
	       &                   &      & 0.03  & 0.18 &      &  0.14 &  &0.009, 0.114  \\
	       &                   &      &       &      &      &       &   &  \\
KKs57          & 134138.1$-$423455 & $-$  & 24.10 & 1.37$\pm$0.033 & 0.18 & 27.97 & 3.93 & $-1.630$ \\
	       &                   &      & 0.05  & 0.19 &      &  0.15 &  &0.062, 0.136  \\
CenN           & 134809.2$-$473354 & $-$  & 24.10 & 1.52$\pm$0.012 & 0.27 & 27.88 & 3.77 & $-1.739$ \\
	       &                   &      & 0.04  & 0.13 &      &  0.15 &  &0.035, 0.143  \\
HIPASS1348--37 & 134833.9$-$375803 &  347 & 24.90 & 1.25$\pm$0.014 & 0.15 & 28.80 & 5.75 & $-2.519$ \\
	       &                   &      & 0.19  & 0.16 &      &  0.24 & &0.065, 0.186  \\
E383--87       & 134918.8$-$360341 &  108 & 23.78 & $-$            & 0.14 & 27.69 & 3.45 & $-$ \\
	       &                   &      & $-$   & $-$  &      &   $-$ & $-$   \\
\hline
\end{tabular}
\end{table}

\hoffset=-1cm
\begin{table}
\tablenum{1}
\caption{--Continued}
\medskip
\begin{tabular}{lcccccccc}
\hline
\hline
Name & RA(J2000.0)Dec & $V_{LG}$ & $I_{TRGB}$ & (V-I)$_{-3.5}$ & $A_I$ & $\mu_0$  & D & [Fe/H] \\
     &               &          & $\sigma_{TRGB}$ & $\sigma_{V-I}$ &  & $\sigma_{\mu_0}$ & & $\sigma_{[Fe/H]}$ \\  
\hline
HIPASS1351--47 & 135122.0$-$470000 &  292 & 25.02 & 1.36$\pm$0.014 & 0.28 & 28.79 & 5.73 & $-2.381$ \\
	       &                   &      & 0.22  & 0.20 &      &  0.26 &  &0.077, 0.179  \\
	       &                   &      &       &      &      &       &     &  \\
E384-016       & 135701.6$-$352002 &  350 & 24.37 & 1.46$\pm$0.005 & 0.14 & 28.28 & 4.53 & $-1.622$ \\
	       &                   &      & 0.03  & 0.14 &      &  0.14 &  &0.015, 0.135  \\
E223--09       & 150108.5$-$481733 &  387 & 25.51 & 1.69$\pm$0.004 & 0.50 & 29.06 & 6.49 & $-1.600$ \\
	       &                   &      & 0.07  & 0.32 &      &  0.16 &  &0.011, 0.134  \\
E274--01       & 151413.5$-$464845 &  335 & 23.90 & 1.81$\pm$0.005 & 0.50 & 27.45 & 3.09 & $-1.217$ \\
	       &                   &      & 0.05  & 0.35 &      &  0.16 &  &0.008, 0.105  \\
E137--18       & 162059.3$-$602915 &  421 & 25.45 & 1.60$\pm$0.005 & 0.47 & 29.03 & 6.40 & $-1.961$ \\
               &                   &      & 0.06  & 0.36 &      &  0.16 &  &0.017, 0.156 \\
\hline
\end{tabular}
\end{table}

\begin{deluxetable}{lcrrrrrrlll}
\tablewidth{0pc}
\tablecaption{Galaxies in and around the Cen~A/M~83 complex.}
\tablehead{
\colhead{Galaxy name}&
\colhead{RA(J2000.0)Dec}&
\colhead{$T$}&
\colhead{$\Theta$}&
\multicolumn{2}{c}{$V_{LG}\pm\sigma_V$}&
\multicolumn{2}{c}{$D\pm\sigma_D$}&
\colhead{Meth}&
\colhead{Ref}&
\colhead{Note}\\
\colhead{(1)}&
\colhead{(2)}&
\colhead{(3)}&
\colhead{(4)}&
\multicolumn{2}{c}{(5)}&
\multicolumn{2}{c}{(6)}&
\colhead{(7)}&
\colhead{(8)}&
\colhead{(9)}}
\startdata
E059--01      &  073119.3$-$681110 &  9 & $-$1.5 & 245 &  5 & 4.57 & 0.36 & rgb & K06a &\\
N2915         &  092611.5$-$763735 & 10 & $-$1.3 & 184 &  5 & 3.78 & 0.43 & rgb & CNG &\\
SexB,DDO70    &  100000.1$+$051956 & 10 & $-$0.7 & 111 &  1 & 1.36 & 0.07 & rgb & CNG &\\
N3109         &  100307.2$-$260936 &  9 & $-$0.1 & 110 &  1 & 1.33 & 0.08 & rgb & CNG &\\
Antlia,P29194 &  100404.0$-$271955 & 10 & $ $2.3 &  66 &  0 & 1.28 & 0.13 & rgb & T06 &\\
SexA,DDO75    &  101100.8$-$044134 & 10 & $-$0.6 &  94 &  1 & 1.32 & 0.04 & cep & CNG &\\
E376--16       &  104327.1$-$370233 & 10 & $ $    & 364 &  3 & 5.0  &      & h   & Ko04& \\
P32250        &  104741.9$-$385115 &  7 & $-$1.0 & 406 &  7 & 5.8  &      & h   & CNG & \\
E215--09,KKs40 &  105730.2$-$481044 & 10 & $-$0.9 & 290 &  2 & 5.25 & 0.41 & rgb & K06b & \\
N3621         &  111816.1$-$324842 &  7 & $-$1.9 & 436 &  5 & 6.70 & 0.47 & cep & CNG & \\
HIPASS        &  113311.0$-$325743 & 10 & $ $    & 414 &  5 & 5.8  &      & h   & M04&  \\
E320--14,KKs44 &  113753.4$-$391314 & 10 & $-$1.2 & 362 &  5 & 6.08 & 0.65 & rgb & K06b &  \\
E379--07,KK112 &  115443.0$-$333329 & 10 & $-$1.3 & 363 &  5 & 5.22 & 0.52 & rgb & CNG &  \\
E379--24       &  120456.7$-$354435 & 10 & $ $    & 354 &  5 & 4.9  &      & h   & M04&  \\
E321--014      &  121349.6$-$381353 & 10 & $-$0.3 & 337 &  5 & 3.19 & 0.26 & rgb & CNG &  \\
I3104,E020--04 &  121846.1$-$794334 &  9 & $-$0.5 & 171 &  5 & 2.27 & 0.19 & rgb & CNG &  \\
KKs51         &  124421.5$-$425623 & $-$3 & $ $0.7 &     &    & 3.6  &      & mem & CNG & C  \\
E381--018      &  124442.7$-$355800 & 10 & $-$0.6 & 365 &  1 & 5.32 & 0.51 & rgb & K06b &   \\
E381--20       &  124600.4$-$335017 & 10 & $-$0.3 & 332 &  6 & 5.44 & 0.37 & rgb & K06b &   \\
HIPASS J1247--77& 124732.6$-$773501 & 10 & $-$1.0 & 155 &  3 & 3.16 & 0.25 & rgb & K06a &   \\
E443--09,KK170 &  125453.6$-$282027 & 10 & $-$0.9 & 410 &  1 & 5.97 & 0.46 & rgb & K06b &   \\
E219--010      &  125609.6$-$500838 & $-$3 & $ $0.1 &     &    & 4.28 &      & sbf & CNG & C \\
GR8,DDO155    &  125840.4$+$141303 & 10 & $-$1.2 & 136 &  3 & 2.10 & 0.34 & rgb & CNG &   \\
UA319         &  130214.4$-$171415 &  9 & $ $2.1 & 547 &  5 & 7.6  &      & h   & CNG &   \\
DDO161        &  130316.8$-$172523 &  8 & $ $1.4 & 545 &  5 & 7.6  &      & h   & CNG &   \\
E269--37,KK179 &  130333.6$-$463503 & $-$3 & $ $1.6 &     &    & 3.48 & 0.35 & rgb & CNG & C \\
KK182,Cen6    &  130502.9$-$400458 & 10 & $-$0.5 & 360 &  3 & 5.78 & 0.42 & rgb & K06b &   \\
N4945         &  130526.1$-$492816 &  6 & $ $0.7 & 296 &  5 & 3.82 & 0.31 & rgb & G05 & C \\
P45628        &  130936.6$-$270826 & 10 & $-$0.6 & 470 & 32 & 6.5  &      & h   & CNG &   \\
E269--058      &  131032.9$-$465927 & 10 & $ $1.9 & 135 &  3 & 3.80 & 0.29 & rgb & K06b & C  \\
KKs53,Cen7    &  131114.2$-$385422 & $-$3 & $ $1.2 &     &    & 3.6  &      & mem & CNG & C  \\
KK189         &  131245.0$-$414955 & $-$3 & $ $2.0 &     &    & 4.42 & 0.33 & rgb & K06b & C  \\
E269--66,KK190 &  131309.2$-$445324 & $-$1 & $ $1.7 & 528 & 31 & 3.82 & 0.26 & rgb & K06b & C  \\
N5068         &  131855.3$-$210221 &  6 & $-$1.4 & 473 &  5 & 6.6  &      & h   & CNG &     \\
KK195         &  132108.2$-$313147 & 10 & $ $0.0 & 338 &  5 & 5.22 & 0.52 & rgb & CNG & M   \\
KKs54         &  132132.4$-$315311 & $-$3 & $ $1.0 &     &    & 4.6  &      & mem & CNG & M    \\
KK196         &  132147.1$-$450348 & 10 & $ $2.2 & 490 &  5 & 3.98 & 0.29 & rgb & K06b & C   \\
N5102         &  132157.8$-$363747 &  1 & $ $0.7 & 230 &  7 & 3.40 & 0.39 & rgb & CNG & C   \\
KK197         &  132201.8$-$423208 & $-$3 & $ $3.0 &     &    & 3.87 & 0.27 & rgb & K06b & C   \\
KKs55         &  132212.4$-$424351 & $-$3 & $ $3.1 &     &    & 3.94 & 0.27 & rgb & K06b & C   \\
KK198         &  132256.1$-$333403 & $-$3 & $ $0.8 &     &    & 4.6  &      & mem & CNG & M    \\
KK200         &  132436.0$-$305820 &  9 & $ $1.2 & 264 &  1 & 4.63 & 0.46 & rgb & CNG & M    \\
N5128,Cen A   &  132528.9$-$430100 & $-$2 & $ $0.6 & 301 &  5 & 3.77 & 0.38 & rgb & R04 & C    \\
I4247,E444--34 &  132644.4$-$302145 & 10 & $ $1.5 & 195 & 10 & 4.97 & 0.49 & rgb & K06b & M    \\
KK203         &  132728.1$-$452109 & $-$3 & $ $2.1 &     &    & 3.6  &      & mem & CNG &  C    \\
E324--24       &  132737.4$-$412850 & 10 & $ $2.4 & 270 &  6 & 3.73 & 0.43 & rgb & CNG & C   \\
P170257       &  132921.0$-$211045 & 10 & $ $0.2 & 457 & 29 & 6.3  &      & h   & CNG &     \\
N5206         &  133343.9$-$480904 & $-$3 & $ $1.1 & 322 & 10 & 3.47 & 0.28 & rgb & Sh06& C   \\
E270-17,RFGC2603 &  133447.3$-$453251 &  8   & $ $1.0 & 583 & 2  & 4.3 & 0.8 &   tf  & K06b & C \\
E444--78,UA365 &  133630.8$-$291411 & 10 & $ $2.1 & 363 &  1 & 5.25 & 0.43 & rgb & K06b & M   \\
KK208         &  133635.5$-$293415 & $-$3 & 1.6 &     &    & 4.68 & 0.46 & rgb & CNG & M   \\
N5236, M83    &  133700.1$-$295204 &  5 & $ $0.8 & 304 &  4 & 5.16 & 0.41 & rgb & K06b & M   \\
DEEP J1337--33 &  133700.6$-$332147 & 10 & $ $1.2 & 371 &  5 & 4.51 & 0.45 & rgb & CNG & M   \\
E444--84       &  133720.2$-$280246 & 10 & $ $1.7 & 380 &  4 & 4.61 & 0.46 & rgb & CNG & M   \\
HIPASS J1337--39& 133725.1$-$395352 & 10 & $-$0.3 & 258 &  5 & 4.90 & 0.49 & rgb & CNG &      \\
N5237         &  133738.9$-$425051 & $-$3 & $ $2.1 & 126 &  5 & 3.40 & 0.23 & rgb & K06b & C   \\
N5253         &  133955.8$-$313824 &  8 & $ $0.0 & 190 &  4 & 3.60 & 0.20 & rgb & Sa04 & C    \\
I4316,E445--06 &  134018.1$-$285340 & 10 & $ $2.4 & 382 & 12 & 4.41 & 0.44 & rgb & CNG & M    \\
N5264         &  134137.0$-$295450 & 10 & $ $2.6 & 268 &  3 & 4.53 & 0.45 & rgb & CNG & M   \\
KKs57         &  134138.1$-$423455 & $-$3 & $ $1.8 &     &    & 3.93 & 0.28 & rgb & K06b & C   \\
KK211         &  134205.6$-$451218 & $-$5 & $ $1.5 & 340 & 23 & 3.58 & 0.36 & rgb & CNG & C   \\
KK213         &  134335.8$-$434609 & $-$3 & $ $1.7 &     &    & 3.63 & 0.36 & rgb & CNG & C   \\
E325--11       &  134500.8$-$415132 & 10 & $ $1.1 & 307 &  4 & 3.40 & 0.39 & rgb & CNG & C   \\
KKs58         &  134600.8$-$361944 & $-$3 & $ $0.6 &     &    & 3.6  &      & mem & CNG & C    \\
KK217         &  134617.2$-$454105 & $-$3 & $ $1.1 &     &    & 3.84 & 0.38 & rgb & CNG & C   \\
KK218         &  134639.5$-$295845 & $-$3 & $ $1.6 &     &    & 4.6  &      & mem & CNG & M    \\
E174--01,KKs59 &  134757.7$-$532104 & 10 & $-$1.5 & 432 &  6 & 6.0  &      & h   & CNG &      \\
CenN          &  134809.2$-$473354 & $-$3 & $ $0.9 &     &    & 3.77 & 0.26 & rgb & K06b & C    \\
HIPASS J1348--37& 134833.9$-$375803 & 10 & $-$1.2 & 347 &  3 & 5.75 & 0.66 & rgb & K06b &      \\
KK221         &  134846.4$-$465949 & $-$3 & $ $0.6 & 270 & 33 & 3.98 & 0.40 & rgb & CNG & C    \\
E383--87       &  134918.8$-$360341 &  8 & $ $0.5 & 108 &  1 & 3.45 & 0.27 & rgb & K06b & C   \\
HIPASS J1351--47& 135122.0$-$470000 & 10 & $-$1.1 & 292 &  3 & 5.73 & 0.73 & rgb & K06b &     \\
KKH86         &  135433.6$+$041435 & 10 & $-$1.5 & 209 &  3 & 2.61 & 0.16 & rgb & CNG &     \\
E384--016      &  135701.6$-$352002 & 10 & $-$0.3 & 350 & 32 & 4.53 & 0.31 & rgb & K06b &     \\
N5408         &  140321.5$-$412235 & 10 & $-$0.5 & 288 &  7 & 4.81 & 0.48 & rgb & CNG &     \\
Circinus,E97--13& 141309.3$-$652021 &  3 & $-$0.7 & 204 & 10 & 2.82 & 0.28 & tfi & K06b &     \\
DDO187        &  141556.5$+$230319 & 10 & $-$1.3 & 172 &  4 & 2.28 & 0.23 & rgb & T06 &     \\
P51659        &  142803.7$-$461806 & 10 & $ $0.1 & 175 &  2 & 3.58 & 0.33 & rgb & CNG & C   \\
E222-10       &  143503.0$-$492518 & 10 & $-$1.4 & 405 &  5 & 5.8  &      & h   & CNG &      \\
HIPASS        &  144127.0$-$624155 & 10 & $ $    & 441 &  3 & 6.1  &      & h   & M04&      \\
E272--25       &  144325.5$-$444219 &  8 & $-$1.5 & 422 & 10 & 5.9  &      & h   & CNG &      \\
E223--09       &  150108.5$-$481733 & 10 & $-$0.8 & 387 &  6 & 6.49 & 0.51 & rgb & K06b &      \\
E274--01       &  151413.5$-$464845 &  7 & $-$0.8 & 335 &  5 & 3.09 & 0.23 & rgb & K06b &     \\
HIPASS J1526--51& 152617.0$-$511004 &    & $ $    & 416 &    & 5.8  &      & h   & M04&     \\
E137--18       &  162059.3$-$602915 &  9 & $-$1.8 & 420 &  3 & 6.40 & 0.48 & rgb & K06b &     \\
I4662,E102--14 &  174706.3$-$643825 &  9 & $-$0.9 & 145 &  4 & 2.44 & 0.19 & rgb & K06a &     \\
Tucana,P69519 &  224149.0$-$642512 & $-$2 & $-$0.1 &   9 &  2 & 0.88 & 0.09 & rgb & CNG &
\enddata
\end{deluxetable}

\begin{table}
\caption{Total mass estimates for the extended halo of Cen A.}
\vspace{1cm}
\begin{tabular}{|lrcl|}  \hline
 Method                &   Scale &       $M_{tot}$      &    Reference     \\
		       &    kpc   &     $10^{12}M_{\sun}$ & \\
\hline
		       &          &                  &                     \\
Globular clusters      &      45  &      0.8 -- 1.8  &    Woodley 2006    \\
		       &          &                  &                     \\
Planetary Nebulae      &      80  &      0.5 -- 0.6   &   Peng et al. 2004 \\
		       &          &                  &                     \\
Orbital/virial        &     400  &      6.4 -- 8.1   &   present paper    \\
		       &          &                  &                     \\
Turn-over radius with &    1400  &      4.4 -- 7.3   &   present paper    \\
 $\Omega_m = 0.27$    &          &                  &              \\
		       &          &                  &                     \\
\hline
		       &          &                  &                     \\
Typical lensing mass  &     500 &       2.4 -- 5.2   & Hoekstra et al. 2005\\
for given $L$           &        &                  &                   \\
		       &          &                  &                     \\
\hline
\end{tabular}
\end{table}

\end{document}